\documentclass[conference]{IEEEtran}
\IEEEoverridecommandlockouts
\usepackage{cite}
\usepackage{subfigure}
\usepackage{amsmath,amssymb,amsfonts}
\usepackage{algorithmic}
\usepackage{graphicx}
\usepackage{textcomp}
\usepackage{xcolor}
\usepackage{hyperref}
\usepackage{color}
\usepackage{amsthm}
\usepackage{comment}
\usepackage{multirow}
\usepackage{array}
\usepackage{diagbox}
\usepackage{booktabs}
\usepackage{tabularx}
\usepackage{bm}

\def\Tenc{T^{\textrm{enc}}}
\def\Tdec{T^{\textrm{dec}}}
\def\Trec{T^{\textrm{rec}}}
\def\Tcmp{T^{\textrm{cmp}}}
\def\Tsen{T^{\textrm{sen}}}
\def\Tw{T^{\textrm{w}}}

\def\Tc{T^{\textrm{c}}}
\def\Tu{T^{\textrm{u}}}

\def\Nenc{N^{\textrm{enc}}}
\def\Ndec{N^{\textrm{dec}}}
\def\Nrec{N^{\textrm{rec}}}
\def\Ncmp{N^{\textrm{cmp}}}
\def\Nsen{N^{\textrm{sen}}}

\def\mum{\mu^{\textrm{m}}}
\def\mutr{\mu^{\textrm{tr}}}

\def\thetam{\theta^{\textrm{m}}}
\def\murec{\mu^{\textrm{rec}}}
\def\thetarec{\theta^{\textrm{rec}}}
\def\musen{\mu^{\textrm{sen}}}
\def\thetasen{\theta^{\textrm{sen}}}
\def\mucmp{\mu^{\textrm{cmp}}}
\def\thetacmp{\theta^{\textrm{cmp}}}
\def\musum{\mu^{\textrm{sum}}}
\def\thetasum{\theta^{\textrm{sum}}}

\newtheorem{lemma}{Lemma}
\newtheorem{definition}{Definition}
\newtheorem{proposition}{Proposition}

\newtheorem{remark}{Remark}

\ifodd 1
\newcommand{\rev}[1]{\textcolor{blue}{#1}}
\newcommand{\revw}[1]{\textcolor{red}{#1}}

\newcommand{\com}[1]{\textbf{\color{red} \left(Comment: #1\right) }}
\newcommand{\comg}[1]{\textbf{\color{blue} \left(COMMENT: #1\right)}}
\newcommand{\response}[1]{\textbf{\color{blue} \left(RESPONSE: #1\right)}}
\else
\newcommand{\rev}[1]{#1}

\newcommand{\revw}[1]{#1}
\newcommand{\com}[1]{}
\newcommand{\comg}[1]{}
\newcommand{\response}[1]{}
\fi

\def\BibTeX{{\rm B\kern-.05em{\sc i\kern-.025em b}\kern-.08em
    T\kern-.1667em\lower.7ex\hbox{E}\kern-.125emX}}
\begin{document}

\title{CoCoI: Distributed Coded Inference System for Straggler Mitigation \vspace{-4mm}
}

\author{Xing Liu, Chao Huang, and Ming Tang\thanks{Xing Liu and Ming Tang are with the Department of Computer Science and Engineering, Southern University of Science and Technology, Shenzhen, China. Chao Huang is with the School of Computing, Montclair State University, Montclair, New Jersey, USA.}
}

\maketitle

\begin{abstract}
    Convolutional neural networks (CNNs) are widely applied in real-time applications on resource-constrained devices. To accelerate CNN inference, prior works proposed to distribute the inference workload across multiple devices. However, they did not address stragglers and device failures in distributed inference, which is challenging due to the devices' time-varying and possibly unknown computation/communication capacities. To address this, we propose a distributed coded inference system, called CoCoI. 
    It splits the convolutional layers of CNN, considering the data dependency of high-dimensional inputs and outputs, and then adapts coding schemes to generate task redundancy.
    With CoCoI, the inference results can be determined once a subset of devices complete their subtasks, improving robustness against stragglers and failures. To theoretically analyze the tradeoff between redundancy and subtask workload, we formulate an optimal splitting problem to minimize the expected inference latency. Despite its non-convexity, we determine an approximate strategy with minor errors, and prove that CoCoI outperforms uncoded benchmarks. For performance evaluation, we build a testbed with Raspberry Pi 4Bs. The experimental results show that the approximate strategy closely matches the optimal solution. When compared with uncoded benchmarks, CoCoI reduces inference latency by up to 34.2\%  in the presence of stragglers and device failures.
\end{abstract}


\section{Introduction}

Convolutional neural networks (CNNs) have been widely applied to various applications, such as image classification and object detection \cite{zhang2023computer}. To meet the real-time and stable response requirements, inference tasks have been increasingly shifted from the cloud to edge devices in order to reduce the communication overhead \cite{Shuvo2023EfficientAO}.
However, CNN inference is typically computationally intensive, while the computation resources of a single edge device are usually too limited to serve inference requests promptly.
This limitation is particularly evident when handling complex models.
For example, 
as shown in Appendix \ref{app:bottleneck},
the inference of 
VGG16\cite{vgg} and ResNet18\cite{resnet} on a Raspberry Pi 4B takes
50.8s and 89.8s, respectively. Meanwhile, the computation of 2-dimensional (2D) convolutional layers are the primary bottleneck of CNN inference. In the aforementioned example, convolutional layers account for more than 99\% of the total latency.

To address this, distributed edge inference  \cite{Hu2019DADS, yang2023ondemand, modnn, dina, coedge} was proposed to overcome the computation bottleneck on a single edge device by distributing computation workload across multiple devices for collaborative inference.
Some of them considered cloud-edge collaboration \cite{Hu2019DADS, yang2023ondemand}, reducing inference workload on edge by assigning part of the model to the cloud.
Some considered edge-edge collaboration\cite{modnn, dina, coedge}.
Such edge-edge collaboration is feasible due to the local dependency of convolutional layers (i.e., each element of convolution output depends on only a local range of input corresponding to its kernel size). In other words, the convolution in CNN can be split for parallel execution at multiple devices,  achieving acceleration.
Specifically, MoDNN\cite{modnn} splits convolution layers based on device computation capacity. DINA\cite{dina} considers the transmission overhead to optimize the multi-requests of convolutions. CoEdge\cite{coedge} enables devices to request necessary resources from others to reduce synchronization overhead.

Unfortunately, distributed systems often suffer from stochastic straggling issues\cite{bitar2020stochastic,zhu2023heterogeneous}, i.e., some devices spend much longer time in task completion than others, and device failure.
Since distributed inference requires
devices to regularly integrate intermediate results\cite{dina}, 
straggling effect and device failure can slow down the integration and hence the overall inference process. Existing works have worked on such straggling/failure issues by considering
straggler detection (e.g., \cite{said2022optimized,tran2023disco}) or introducing replication-based task redundancy (e.g., \cite{behrouzi2019data,ciucu2021Prac}).
However, these methods may not be robust under highly dynamic scenarios or lead to significant resource waste. 
To deal with these straggling/failure issues, some existing works have proposed coding schemes for general distributed computation. 
For example, Lee \emph{et al.} \cite{Lee2018speeding} proposed to apply MDS codes to accelerate distributed matrix-vector multiplication. Mallick \emph{et al.} \cite{Mallick2022rateless} proposed fountain codes that are more robust to accelerate large-scale matrix multiplication. 
However, although such codes may be feasible for the distributed computation of convolution \cite{coded_convolution, Zhou2022Dynamic}, those existing works (e.g., \cite{Lee2018speeding,Mallick2022rateless}) did not consider the complex data dependency brought by the high-dimensional inputs when decomposing the convolutions.

\begin{figure*}[t]
    \centering
    \includegraphics[height=4.8cm]{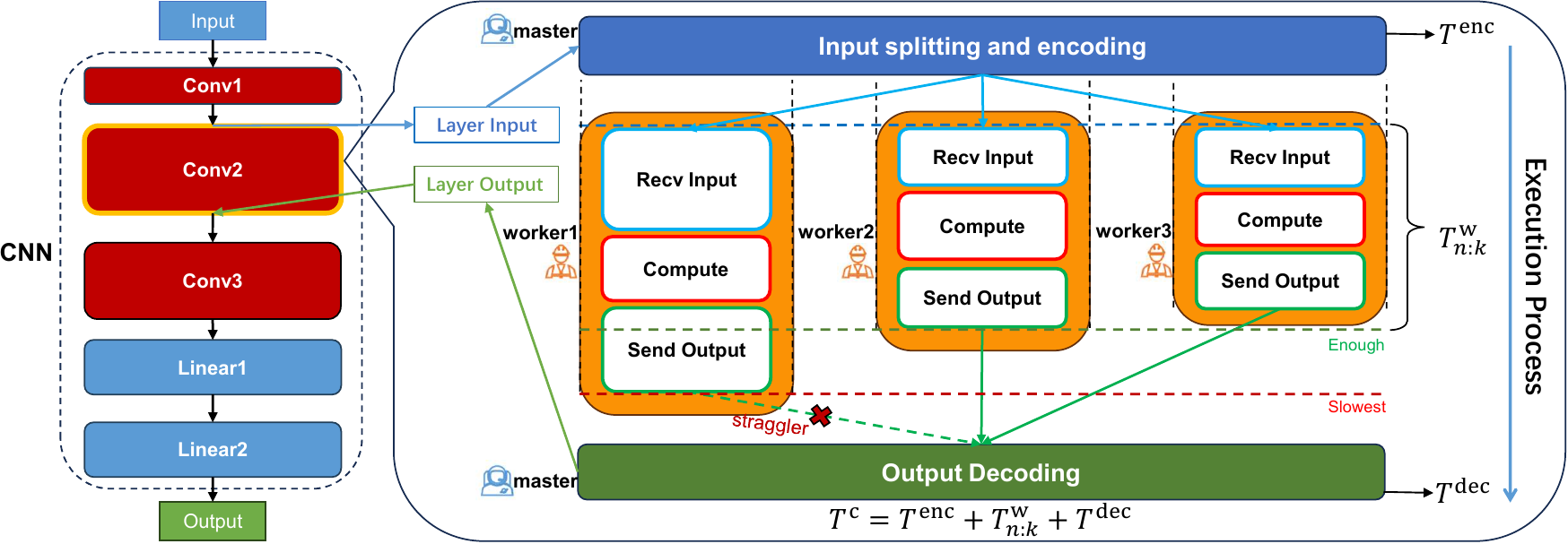}\vspace{-2mm}
    \caption{An illustration of the CoCoI system with $n=3$ and $k=2$ and the workflow of our distributed coded inference approach applied to layer \textit{Conv2}. 
    The master first splits the input of layer \textit{Conv2} and extends the input partitions to $n=3$ encoded input partitions, each corresponding to an encoded subtask. Then, the workers receive their assigned inputs, perform execution, and  send their outputs to the master. 
    Once $k=2$ outputs have been received, the master starts decoding to obtain the layer output. Here, $\Tenc, \Tw_{n:k}, \Tdec$ denote the encoding, execution, and decoding latency (see Section \ref{sec:problme} for details), respectively.} 
    \label{fig:system}
\end{figure*}

To address straggling and device failure issues in distributed CNN inference, we propose \underline{Co}ded \underline{Co}operative \underline{I}nference (CoCoI), a distributed coded inference system that introduces task redundancy to  improve system robustness against straggler and device failure. 
Through task splitting and encoding design, once a subset of devices have accomplished their assigned executions on CNN inference (conditioning on the number of devices which accomplished), their outputs are sufficient to be decoded to obtain the final inference result, thus efficiently eliminating straggler/failure issues.
Note that designing a distributed coded computation method for general CNNs
can be challenging due to the complex data dependency of high-dimensional inputs. Meanwhile, formulating and analyzing the associated optimal splitting problem is non-trivial due to the randomness involved in straggling effect and its non-convexity.
We summarize our contributions as follows:
\begin{itemize}
    \item \textbf{Distributed Coded Inference System}:
    We propose a distributed coded computation system for CNN inference, called CoCoI. Through task splitting and encoding/decoding design, this system enables devices to cooperatively execute computational-intensive layers of CNN and mitigate the straggling effect and device failure, while introducing only minor encoding and decoding overhead. 
    \item \textbf{Optimal Splitting and Performance Analysis}:
    We take into account the straggling effect in CoCoI and formulate an optimal splitting problem. Although its objective function lacks an explicit form and is shown to be non-convex, we transform the problem into an approximate convex problem, with which we determine an approximate optimal solution. We further analyze how system parameters affect the approximate solution. Moreover, based on the approximate problem, we theoretically prove that CoCoI always achieves a lower inference latency than uncoded scheme in the presence of  stragglers.
    \item \textbf{Performance Evaluation on Testbed}: We evaluate our CoCoI system on a real-world testbed with Raspberry-Pi 4Bs. 
    Experimental results show that the performance gap between our approximate optimal splitting strategy and the optimal solution is  less than 3.3\%. When compared to uncoded and replication-based benchmarks, CoCoI system can reduce the inference latency by up to 34.2\% in the presence of straggler or device failure.

\end{itemize}

The rest of this paper is organized as follows.
Section \ref{sec:model} 
presents our system. 
Section \ref{sec:problme} 
formulates the optimal splitting problem.  
In Section \ref{sec:analysis}, we analyze the optimal strategy and its performance. Section \ref{sec:experiment} shows our testbed and experimental results. Section \ref{sec:conclusion} concludes this work. 

\section{System Design}\label{sec:model}


\par In this section, we first propose the CoCoI system. Then, we present our distributed coded inference approach, which is an important component of the CoCoI system. 

\subsection{CoCoI System Overview}

In CoCoI system, there is a master device\footnote{The master can be served by the device which initiates the inference task. Since this device has inference request, it will stay in the system during  inference. Meanwhile, since the workload at the master is light, even if the master sometimes fails, re-execution can be completed within a short period of time (e.g., in hundreds of milliseconds). Thus, we do not introduce schemes to address the device failure of the master. This setting is consistent with  existing  works on coded computation \cite{Zhou2022Dynamic, fang2023latency}.}  and $n$ worker devices. 
The master tracks the CNN inference process, distributes computation tasks to workers, and performs the related encoding and decoding operations. The workers are responsible for task execution. 
To complete a CNN inference task, CNN layers should be executed in a specific order.
We refer to the execution of each layer as a \emph{computation task}. Based on computational complexity (see Appendix \ref{app:bottleneck}),
we classify the
layers into two types: 
\begin{itemize}
    \item Type-1 task (high-complexity): Most 2D convolutional layers, which is the bottleneck of CNN inference. 
    \item Type-2 task (low-complexity): Linear, activation, and some light-weight 2D convolutional layers.
\end{itemize}


The CoCoI adopts a distributed coded inference approach (to be proposed in Section \ref{subsec:distributed}) to execute the high-complexity type-1 tasks in order to eliminate the inference bottleneck.
This process contains four phases: input splitting phase for input feature map partitioning, encoding phase to generate task redundancy, execution phase for distributed task execution, and decoding phase for recovering the task output.  
The master device executes those low-complexity type-2 tasks locally to mitigate the communication overhead resulting from intermediate computation  result exchange. 



An illustration of  CoCoI system is shown in Fig. \ref{fig:system}.
The left part of Fig. \ref{fig:system} shows the layers of a CNN, 
where  convolutional layers Conv1, Conv2, and Conv3 are executed in a distributed manner.
The right part of Fig. \ref{fig:system} shows the workflow of the distributed coded inference approach for a type-1 task. 

\begin{figure*}[t]
    \centering
    \includegraphics[width=0.75\textwidth]{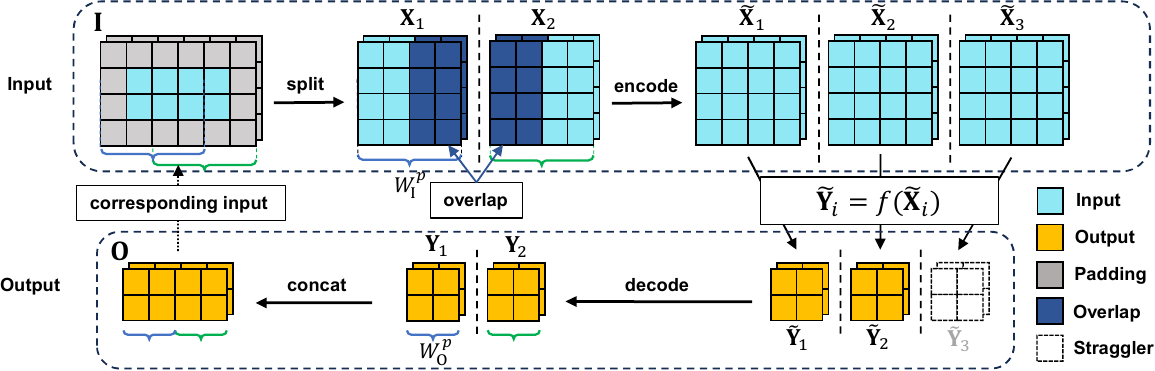}\vspace{-3mm}
    \caption{An illustration of input splitting, encoding, and output decoding for distributed coded convolution with a $3\times3$ kernel and $\text{stride}=1$. Here, an $(n,k)$-MDS code ($n=3$ and $k=2$) is used to encode the input partitions and to decode the computation task result based on the encoded outputs.
    }
    \label{inputoutput}
\end{figure*}
\subsection{Distributed Coded Inference Approach}\label{subsec:distributed}
 In the distributed coded inference approach, 
we partition the type-1 convolutional layers
and perform  data transform on the input feature map to ensure distributed inference. 
Meanwhile, we propose to apply MDS code to introduce task redundancy in order to mitigate straggler/failure issues. Although other codes (e.g., fountain codes) can also be modified to support the encoding and decoding design, they are not suitable to CoCoI due to their overhead and reduced level of successful decoding rate (see Section \ref{sec:experiment} for experimental results).

Specifically, a 2D convolutional layer in CNNs mainly has five configuration parameters, including \textit{in\_channels}, \textit{out\_channels} (denoted by $C_\textsc{I},C_\textsc{O}$), \textit{kernel\_size}, \textit{stride}, and \textit{paddings}.
We use $K_W$ and $S_W$ to represent the values of \textit{kernel\_size} and \textit{stride} on the width dimension, respectively. Meanwhile, we set the \textit{kernel\_size} on both width and height dimensions to be the same, i.e., $K_W$.
The original input feature map is firstly padded with the \textit{paddings} (the gray area surrounding the input feature map in Fig. \ref{inputoutput}). Then, the 4D padded input (denoted by $\mathbf{I}$) has a shape of $(B,C_\textsc{I},H_\textsc{I},W_\textsc{I})$, where $B$ is the batch size and $H_\textsc{I},W_\textsc{I}$ denote the height and width of $\mathbf{I}$, respectively. Similarly, let $(B,C_\textsc{O},H_\textsc{O},W_\textsc{O})$ denote the shape of output $\mathbf{O}$. Here we set $B=1$ considering sparse inference requests on edge. Without loss of generality, we assume $W_\textsc{I}\geq H_\textsc{I}$. In this case, we split the feature map in the width dimension (rather than the height dimension) as it can reduce the overlap of input partitions. Based on the principle of 2D convolution, $W_\textsc{O}$ can be determined by $W_\textsc{O}=\left\lfloor {(W_\textsc{I} - K_W+1)}/{S_W}+1\right\rfloor$ (and $H_\textsc{O}$ similarly).
Key parameters are summarized in Table \ref{tab:params}.

\subsubsection{Input Splitting Phase}

The master splits the input feature map into $k$ pieces, corresponding to the input of $k$ \emph{source subtasks}. The input  partition is determined by computing the target output $\mathbf{O}$ of each source task, and the respective partitions is obtained based on the dependency between input and output.
To fulfill the requirement of MDS codes, $\mathbf{O}$ is split into $k$ pieces $\mathbf{Y}_1,\dots,\mathbf{Y}_k$ on the width dimension in equal, each with a shape of $(B,C_\textsc{O},H_\textsc{O},W_\textsc{O}^p(k))$ and an output range of $(a_{\textsc{O}i}, b_{\textsc{O}i})$ for $i\in[k]$ (i.e., a segment of $\mathbf{O}$ from width index $a_{\textsc{O}i}$ to $b_{\textsc{O}i}-1$), where $W^p_\textsc{O}(k) = b_{\textsc{O}i} - a_{\textsc{O}i} = \lfloor {W_\textsc{O}}/{k}\rfloor$\footnote{For the cases where $W_\textsc{O}$ is not divisible by $k$, 
the master can keep the subtask associated with the remaining part of the output feature map with width $\text{mod}(W_\textsc{O},k)$. This subtask is relatively small and has no data transmission latency, thus will not be the bottleneck of the inference process.}.
Consider a piece with an output range of $(a_\textsc{O},b_\textsc{O})$, the width $W_\textsc{I}^p(k)$ and range $(a_\textsc{I},b_\textsc{I})$ of its  input partition is as follows:
\begin{equation}
W_\textsc{I}^p(k)=K_W+(W_\textsc{O}^p(k)-1)\times S_W,
\end{equation}
\begin{equation}
    a_\textsc{I}={a_\textsc{O}}\times S_W, b_\textsc{I}=({b_\textsc{O}}-1)\times S_W+K_W.
\end{equation}
The master splits $\mathbf{I}$ based on the input ranges derived above to obtain input partitions $\mathbf{X}_1,\dots,\mathbf{X}_k$ 
to ensure correct input-output correspondence for each source subtask, i.e., $\mathbf{Y}_i=f(\mathbf{X}_i), i\in[k]$. 
Note that the adjacent input partitions 
may overlap
(see Fig. \ref{inputoutput}), therefore $k\times W^p_\textsc{I}(k)\geq W_\textsc{I}$.

\subsubsection{Encoding Phase}
The master generates the inputs for $n$ \emph{computation subtaks} based on the input of $k$ source subtasks to introduce redundancy. Specifically, $\mathbf{X}_1,\dots,\mathbf{X}_k$ are first flatten\footnote{An operation to transform a high-dimensional array data to an one-dimensional vector, detailed description can be found in \cite{pytorch_docs}.} to vectors $\bm{x}_1,\dots,\bm{x}_k$ (i.e.,  $\bm{x}_i=\text{flatten}(\mathbf{X}_i)$) and concatenated to an input matrix of shape $(k,B\times C_\textsc{I}\times H_\textsc{I}\times W_\textsc{I}^p(k))$.
Then, an $n\times k$ generation matrix $\bm{G}$ is applied on the input matrix to generate the following encoded input matrix, composed of $n$ encoded input vector $\Tilde{\bm{x}}_1,\dots,\Tilde{\bm{x}}_n$:
\begin{equation}\label{MDS-encoding}
    \bm{G} \begin{bmatrix}
        \bm{x}_1 \\ \vdots \\ \bm{x}_k
    \end{bmatrix}
    =
    \begin{bmatrix}
        g^{k-1}_1 & g^{k-2}_1 & \cdots & g^0_1 \\
        \vdots & \vdots & \ddots & \vdots \\
        g^{k-1}_n & g^{k-2}_n & \cdots & g^0_n
    \end{bmatrix}
    \begin{bmatrix}
        \bm{x}_1 \\ \vdots \\ \bm{x}_k
    \end{bmatrix}
    =
    \begin{bmatrix}
        \Tilde{\bm{x}}_1 \\ \vdots \\ \Tilde{\bm{x}}_n
    \end{bmatrix}.
\end{equation}

Here, $\bm{G}$ should satisfy that every $k$-row submatrix of $\bm{G}$ is invertible to ensure successfully decoding, and Vandermonder matrix is usually used as the generation matrix \cite{Lee2018speeding}. 
Then $\Tilde{\bm{x}}_1,\dots,\Tilde{\bm{x}}_n$ are restored to the original shape as 
$\Tilde{\mathbf{X}}_1,\dots,\Tilde{\mathbf{X}}_n$, serving as the input of $n$ computation subtask for execution.



\subsubsection{Execution Phase}
The master sends the $n$ encoded input partitions $\Tilde{\mathbf{X}}_i$ to $n$ workers. Worker $i\in[n]$ receives the encoded input and feeds $\Tilde{\mathbf{X}}_i$ into the on-processing layer for computation subtask execution with the preloaded weights and obtains encoded output $\Tilde{\mathbf{Y}}_i=f(\Tilde{\mathbf{X}}_i)$ of shape $(B,C_\textsc{O},H_\textsc{O},W^p_\textsc{O}(k))$. Afterwards, $\Tilde{\mathbf{Y}}_i$ is sent back to master.

\subsubsection{Decoding Phase} When the master receives the encoded outputs  from any $k$ workers, it can decode and obtain the final output. Let $\mathcal{S}=\{s_1,\dots,s_k\}\subset[n]$ denotes the index set of the $k$ fastest workers.
Then the $k$ encoded outputs $\Tilde{\mathbf{Y}}_{s_1},\dots,\Tilde{\mathbf{Y}}_{s_k}$ are flatten to vectors $\Tilde{\bm{y}}_{s_1},\dots,\Tilde{\bm{y}}_{s_k}$ and concatenated to an encoded output matrix of shape $(k,B\times C_\textsc{O}\times H_\textsc{O}\times W^p_\textsc{O}(k))$. A $k\times k$ sub-matrix $\bm{G}_\mathcal{S}$
is obtained by sequentially selecting the rows of the $\bm{G}$ based on $\mathcal{S}$. The inverse of $\bm{G}_\mathcal{S}$, denoted by $\bm{G}^{-1}_\mathcal{S}$, is applied on the encoded output matrix to generate the decoded output matrix, composed of $k$ output partition vectors $\bm{y}_1,\dots,\bm{y}_k$. That is, 
\begin{equation}\label{MDS-decoding}
    \begin{bmatrix}
        \bm{y}_1 \\ \vdots \\ \bm{y}_k
    \end{bmatrix}
    \!\!=\!\bm{G}^{-1}_\mathcal{S}\! \begin{bmatrix}
        \Tilde{\bm{y}}_{s_1} \\ \vdots \\ \Tilde{\bm{y}}_{s_k}
    \end{bmatrix}\!\!,
    \text{where}\ \bm{G}_\mathcal{S}\!=\!\begin{bmatrix}
        g^{k-1}_{s_1} & g^{k-2}_{s_1} & \cdots & g^0_{s_1} \\
        \vdots & \vdots & \ddots & \vdots \\
        g^{k-1}_{s_k} & g^{k-2}_{s_k} & \cdots & g^0_{s_k}
    \end{bmatrix}\!\!.
\end{equation}
Vectors $\bm{y}_1,\dots,\bm{y}_k$ are then restored to the original shape and concatenated to obtain $\mathbf{O}=\text{concat}(\mathbf{Y}_1,\dots,\mathbf{Y}_k)$.
With the MDS mechanism, $\mathbf{O}$ can be perfectly restored and remains consistent with that of the original inference, thus keeping the inference quality unchanged.


\section{Optimal Splitting Problem}\label{sec:problme}
In CoCoI system, given a total of $n$ workers, the choice of $k$ (i.e., how the task is split) introduces a tradeoff between the computation workload  and redundancy. Specifically, a larger $k$ leads to a reduced  size of each input partition and thus a reduced workload of each subtask. On the other hand, a smaller $k$ leads to an increased  degree of redundancy among workers (i.e., a larger $r\triangleq n-k$), which reduces the execution latency considering the $k$ fastest workers, and slightly reduces the encoding and decoding latency. 
To investigate this tradeoff, 
we formulate an optimal splitting problem to minimize the inference latency. When compared with existing works on distributed inference (e.g.,\cite{dina, coedge}), we characterize the straggling/failure effect by modeling the latency of phases as random variables. When compared with existing modeling of stragglers in coded convolution (e.g., \cite{coded_convolution, Zhou2022Dynamic}), we take into account the modeling of encoding and decoding latency and CNN task-specific execution and transmission latency.

In the following, we first present the overall latency model (of a computation task). Then, we discuss the detailed latency modeling in each phase. Finally, we formulate the problem.

\subsection{Overall Latency Model}
Given the choice of splitting strategy $k$, the overall latency of a computation task consists of three terms: (i) the latency of input splitting and encoding $\Tenc(k)$, (ii) the latency of input/output transmission and execution $\Tw_{n:k}(k)$, and (iii) the latency of  decoding $\Tdec(k)$. That is:
\begin{equation}
\Tc(k)=\Tenc(k)+\Tw_{n:k}(k)+\Tdec(k),
\end{equation}
where $\Tenc(k)$,  $\Tw_{n:k}(k)$, and $\Tdec(k)$ are random variables.

We now present how random variable $\Tw_{n:k}(k)$ is computed. Specifically, suppose worker $i\in[n]$ receives the assigned input with latency $\Trec_i(k)$. It then executes subtask with latency $\Tcmp_i(k)$ and sends the output to the master with latency $\Tsen_i(k)$. Note that $\Trec_i(k)$, $\Tcmp_i(k)$, and $\Tsen_i(k)$ are random variables. Thus, the latency that worker $i$ takes to finish the transmission and execution operations mentioned in  (ii), denoted by $\Tw_i(k)$, can be expressed as follows:
\begin{equation}
\Tw_i(k)=\Trec_i(k)+\Tcmp_i(k)+\Tsen_i(k), ~i\in[n]. 
\end{equation}
Recall that master can decode the final result once receiving $k$ encoded outputs from workers.
Thus, the latency in (ii) is equal to the latency $\Tw_i(k)$ of the $k$-th fastest workers, i.e., $\Tw_{n:k}(k)$. Note that $\Tw_{n:k}(k)$ denotes the $k$-th order statistics of $n$ random variables in set $\{\Tw_1(k),\Tw_2(k),...,\Tw_n(k)\}$, i.e., the $k$-th smallest value among these $n$ random variables. 


Suppose these random variables $\Trec_i(k)$, $\Tcmp_i(k)$, $\Tsen_i(k)$, $\Tenc(k)$, and $\Tdec(k)$ follow a shift-exponential distribution. This is reasonable because an exponential distribution (with or without a shift) usually indicates the time interval between independent events and is widely adopted in modeling the latency of computation tasks in coded computation \cite{marshall1967multivariate}.
Meanwhile, experiments on our testbed (see Appendix \ref{exponential_fitting})
and other research\cite{hcmm, sharedcoded} demonstrate that the latency of computation or wireless transmission can be well-fitted by the shift-exponential distribution. 
\begin{definition}[Shift-Exponential Distribution]\label{def:exp} Consider a random variable $T$ follows a {shift-exponential distribution} defined by a straggler parameter $\mu$, a shift coefficient $\theta$, and a scaling parameter $N$. The CDF of the  random variable $T$  is given by 
\begin{equation}
    F_{\textsc{se}}(t; \mu, \theta, N) \triangleq 1-e^{-\frac{\mu}{N}(t-N\theta)}, t\geq N\theta.
\end{equation}
\end{definition}
In this distributed inference system, 
a smaller straggler parameter $\mu$ implies a stronger straggling effect. The shift coefficient $\theta$ indicates the minimum completion time of the corresponding operation. The scaling parameter $N$ describes the scale of the specific operation, e.g.,  the size of transmission or computation. If a device takes a very long time (e.g., longer than a pre-defined timeout threshold) on a certain operation, we regard the operation as failed. Thus, the scenario of device failure can be captured by shift-exponential distribution. 

 

\subsection{Detailed Latency Model}\label{latency_formulation}
Let  $\mum$, $\mucmp$, $\murec$, and $\musen$ denote the  straggling parameters of the computation at master, computation, input receiving, and output sending at workers, respectively. Let  $\thetam$, $\thetacmp$, $\thetarec$, and $\thetasen$ denote the associated minimal completion time. In practical systems, these straggling and shift coefficients can be estimated by prior test and fitting. In the following, we model the latency of each phase by specifying the scaling parameter $N$ in Definition \ref{def:exp}, considering the specific partition and coding schemes in CoCoI system.

\subsubsection{Latency in Input Splitting and Encoding}\label{subsec:encode}
The master applies the $n\times k$ generation matrix $\bm{G}$ on the input matrix of shape $(k,B\times C_\textsc{I}\times H_\textsc{I}\times W_\textsc{I}^p(k))$ to generate encoded inputs.
Thus, the number of floating point operations (FLOPs) in this phase, denoted by $\Nenc(k)$, can be derived by:
\begin{equation}
\Nenc(k)=2k\times n\times B\times C_\textsc{I}\times H_\textsc{I}\times W_\textsc{I}^p(k).
\end{equation}
The encoding latency $\Tenc(k)$   follows 
$F_{\textsc{se}}\left(t; \mum, \thetam, \Nenc(k)\right)$. 

\subsubsection{Latency in Input/Output Transmission and Execution} 
To execute the subtask, each worker 
performs a sliding dot-product using a kernel of shape $(B,C_\textsc{I},K_W,K_W)$ across the received input feature map and generates an output feature map with shape $(B,C_\textsc{O},H_\textsc{O},W_\textsc{O}^p(k))$. Then, the FLOPs $\Ncmp(k)$ of each subtask on worker is given by:
\begin{equation}
\Ncmp(k)=B\times C_\textsc{O}\times H_\textsc{O}\times W^p_\textsc{O}(k)\times 2\times C_\textsc{I}\times K_W^2.
\end{equation}
Thus, the computation latency  $\Tcmp_i$ on worker $i\in[n]$ follows a shift-exponential distribution   $F_{\textsc{se}}(t;\mucmp,\thetacmp,\Ncmp(k))$. 

The transmission size for the input distributing $\Nrec(k)$ by the master and output forwarding $\Nsen(k)$ by each worker can be formulated by the size of the input and output partitions:
\begin{equation}
\Nrec(k)=4\times B\times C_\textsc{I}\times H_\textsc{I}\times W^p_\textsc{I}(k),
\end{equation}
\begin{equation}
\Nsen(k)=4\times B\times C_\textsc{O}\times H_\textsc{O}\times W^p_\textsc{O}(k).
\end{equation}
The workers share the same  $\Nrec(k)$ and $\Nsen(k)$ due to equal-size subtask partition. The latency $\Trec_i(k)$ and $\Tsen_i(k)$ of the input/output sending 
of  worker $i\in[n]$ respectively follow  $F_{\textsc{se}}(t;\murec,\thetarec,\Nrec(k))$ and  $F_{\textsc{se}}(t;\musen,\thetasen,\Nsen(k))$. 
\subsubsection{Latency in Decoding}
The master decodes the execution output matrix with shape $(k, B\times C_\textsc{O}\times H_\textsc{O}\times W^p_\textsc{O}(k))$ with the $k\times k$ inverse matrix $\bm{G}^{-1}_\mathcal{S}$. Similarly, the FLOPs of decoding phase $\Ndec(k)$ can be derived by:
\begin{equation}
\Ndec(k)=2k^2\times B\times C_\textsc{O}\times H_\textsc{O}\times W_\textsc{O}^p(k).
\end{equation}
The 
decoding latency $\Tdec(k)$  follows $ F_{\textsc{se}}(t; \mum, \thetam, \Ndec(k))$.

\subsection{Optimal Splitting Problem Formulation}\label{subsec:problem}

Our objective is to find the optimal choice of $k^*$ in order to minimize the expectation of overall latency $\mathbb{E}[\Tc(k)]$: 
\begin{equation}\label{P1}
   k^*=\arg \min_{k\in\{1,2,\cdots,n\}} \quad \mathbb{E}[\Tc(k)].
\end{equation}
\begin{remark}[Challenges of Solving Problem \eqref{P1}]
    It is difficult to represent the objective function as an explicit expression due to the involvement of order statistics. Meanwhile, we empirically validate that the objective function $\mathbb{E}[\Tc(k)]$ is non-convex. Both reasons make solving problem \eqref{P1} challenging.
\end{remark}

\section{Optimal Splitting Analysis}\label{sec:analysis}

In this section, we aim to approximately solve problem \eqref{P1}, with which we investigate  (i) how to determine an approximate optimal strategy $k^{\circ}$, (ii) how system parameters (e.g., $\mucmp,\thetacmp, \mum$) affect the optimal strategy, and (iii) how much is the latency reduction from the CoCoI system. 

\subsection{Approximate Optimal Analysis}
Since the encoding $\Tenc(k)$ and decoding $\Tdec(k)$ latency at the master and the latency  $\Tw_{n:k}(k)$ at workers are independent, the objective function in \eqref{P1} can be represented as 
\begin{equation}\label{eq:obj-eq}
    \mathbb{E}[\Tc(k)]=\mathbb{E}[\Tenc(k)+\Tdec(k)]+\mathbb{E}[\Tw_{n:k}(k)],
\end{equation} 
where the expected sum of the encoding and decoding latency $\mathbb{E}[\Tenc(k)+\Tdec(k)]=(\Nenc(k)+\Ndec(k))\times(\frac{1}{\mum}+\thetam)$. Thus, the main difficulty  of solving problem \eqref{P1} comes from determining the expectation of execution latency $\mathbb{E}[\Tw_{n:k}(k)]$. This is because the execution latency $\Tw_{n:k}(k)$ is the $k$-th order statistics of  $\Tw_i(k)$, where  $\Tw_i(k)$ is  the sum of three random variables $\Trec_i(k)$, $\Tcmp_i(k)$, and $\Tsen_i(k)$. Calculating the expectation of order statistics over a sum of multiple random variables is still an open problem  \cite{david2004order}.



Let $\Trec_{n:k}(k)$, $\Tcmp_{n:k}(k)$, and $\Tsen_{n:k}(k)$ denote the associated $k$-th order statistics (i.e., the $k$ smallest values) among workers $i\in[n]$. We approximate $\mathbb{E}[\Tw_{n:k}(k)]$ as follows: 
\begin{equation}\label{eq:app}
    \mathbb{E}[\Tw_{n:k}(k)]\approx \mathbb{E}[\Trec_{n:k}(k)]+\mathbb{E}[\Tcmp_{n:k}(k)]+\mathbb{E}[\Tsen_{n:k}(k)].
\end{equation}
Based on the expectation of exponential order statistics\cite{david2004order} and relaxing the floor function in $W^p_\textsc{O}(k)=\lfloor{W_\textsc{O}}/{k}\rfloor$, an approximate expected overall latency is given by:
\begin{multline}\label{eq:tc-approx}
\mathbb{E}[\Tc(k)] \approx L(k) \triangleq  (\Nenc(k)+\Ndec(k))\left(\frac{1}{\mum}+\thetam\right)\\
+\thetasum+\musum\ln\frac{n}{n-k}, 
\end{multline}
where $\musum\triangleq {\Nrec(k)}/{\murec}+{\Ncmp(k)}/{\mucmp}+{\Nsen(k)}/{\musen}$  and  $\thetasum\triangleq \Nrec(k)\thetarec+\Ncmp(k)\thetacmp+\Nsen(k)\thetasen$. 

An approximate optimal solution  $k^{\circ}$ to problem \eqref{P1} can be determined by minimizing $L(k)$:
\begin{equation}\label{eq:p2}
    k^{\circ} =\arg  \min_{k\in\{1,2,...,n\}} \quad L(k).
\end{equation}

To understand the approximate optimal solution $k^{\circ}$, we temporarily relax integer $k$ to be a real value and eliminate the case of $n=k$ (i.e., no redundancy), i.e., $k\in[1,n)$. 
\begin{lemma}\label{lem:convex}
When $n\geq 3$, the relaxed problem \eqref{eq:p2} is a convex programming problem under $k\in[1,n)$.
\end{lemma}


The proof of Lemma \ref{lem:convex} is detailed in Appendix \ref{proof_convex}. 
Thus, when compared to $\mathbb{E}[\Tc(k)]$, function $L(k)$ can be represented in an explicit form, and it is a convex function given the integer relaxation considered in Lemma \ref{lem:convex}. To determine $k^\circ$, we can first solve $k'=\arg \min_{k\in[1,n)} L(k)$  using conventional solvers (e.g., CVX). Then, we can obtain $k^\circ\in\{\lfloor k'\rfloor,\lceil k'\rceil\}$ by comparing $L(\lfloor k'\rfloor)$ and $L(\lceil k'\rceil)$.

In Section \ref{sec:experiment} and Appendix \ref{appendix:eval},
we empirically evaluate the gap between the approximate optimal solution $k^\circ$ and the optimal solution $k^*$ and their performance gap. The experimental results show that problem \eqref{eq:p2} has a good approximation for (i) the objective function and (ii) the optimal $k$ under a wide range of values of $\mu$'s and $\theta$'s. In most cases, the difference of $k^*$ and $k^\circ$ does not exceed 1.

\subsection{Impact of System Parameters}


Let $\hat{k}^\circ$ denote the optimal solution to problem \eqref{eq:p2} under the relaxation of $k\in[1,n)$. Based on Lemma \ref{lem:convex}, by checking the first-order condition of $L(k)$,  we analyze the impact of system parameters on $\hat{k}^{\circ}$ (see Appendix \ref{proof_analysis_impact} for proof). 
The analytical results on $\hat{k}^{\circ}$ are consistent with the empirical optimal solution $k^*$ to problem \eqref{P1} (see Appendix \ref{appendix:eval}).
\begin{proposition}[Impact of Straggler and Shift Coefficients]\label{prp:impact}
The approximate optimal $\hat{k}^{\circ}$ increases if  (i) any straggler coefficient $\mucmp$, $\mum$, $\murec$, $\musen$ increases, (ii) any shift coefficient $ \thetacmp$, $\thetarec$,  $\thetasen$ increases, or (iii) shift coefficient  $\thetam$ decreases.
\end{proposition}
As shown in Proposition \ref{prp:impact}, if any straggler coefficient $\mu$ decreases (i.e., a more severe straggling effect),  
then 
the computation task should be partitioned into fewer pieces (i.e., smaller $k$) to introduce more redundancy (i.e., larger $r\triangleq n-k$). If any shift coefficient $\theta$ of workers (i.e., $\thetacmp, \thetarec,\thetasen$) increases (i.e., the minimum completion time increases), then the workload of subtasks become larger, so the computation task should be partitioned into smaller pieces (i.e., a larger $k$) to reduce the workload. If $\frac{1}{\mum}+\thetam$ is larger, the master has a less powerful processing capacity, so $k$ should be decreased to reduce the encoding and decoding latency.

\subsection{Theoretical Performance Improvement}\label{cp2uncoded}
For analytical simplicity, we use the approximate form $L(k)$ in the following analysis. The expected latency under uncoded approach can be regarded as a special case of that under coded approach, i.e., the master can obtain the execution result of each layer when all $n$ workers send back their output. Thus, it can be computed based on \eqref{eq:app} and \cite{david2004order} (see Appendix \ref{appendix:performance}).
To provide a clear insight on comparison, we consider a setting where $W_\textsc{O}\gg k$, which matches the case in practical systems. We omit the encoding and decoding latency in coded approach, as those latency are minor (see Section \ref{sec:experiment}). With these approximation, let $\mathbb{E}[\Tc_m(n,k)]$ and $\mathbb{E}[\Tu_m(n)]$ denote the expected latency of coded and uncoded method, respectively. The proof for the following Propositions \ref{prp:cp2uncoded} and \ref{prop:fail} are given in Appendix \ref{appendix:performance}.


First, we consider the straggler scenario without worker failure. Let  $R \triangleq (4I_W\thetarec\!+\!4O\thetasen\!+\!N_{c}\thetacmp)/(\frac{4I_W}{\murec}+\frac{4O}{\musen}+\frac{\Ncmp_{t}}{\mucmp})$, where $I_W\triangleq C_\textsc{I}H_\textsc{I}W_\textsc{O}S$, $O\triangleq C_\textsc{O}H_\textsc{O}W_\textsc{O}$, and $\Ncmp_{t}\triangleq 2C_\textsc{O}H_\textsc{O}C_iK^2W_\textsc{O}$. Intuitively, a smaller $R$ implies a higher degree of straggling effect. 
\begin{proposition}[Straggler Scenario]\label{prp:cp2uncoded}
When $R\leq1$ and $n\geq10$, there exists a $k^*_{sub}\in(1,n)$ such that $\Delta\triangleq\mathbb{E}[\Tu_m(n)]-\mathbb{E}[\Tc_m(n,k^*_{sub})]>0$.
\end{proposition}

Proposition \ref{prp:cp2uncoded} implies that when the  degree of straggling exceeds a threshold (i.e., $R\leq 1$), our coded inference approach always achieves a lower latency than uncoded method \cite{dina}. 
For example, when $n=20$ and $R=1$, our approach  reduces the latency by around $21\%$.

Second, consider a scenario where one  worker eventually fails. For the uncoded approach, we assume that the worker will signal the master when failure happens, and the master will send the subtask to another device for re-execution. Let $R^{\textrm{cmp}}(n)=\mathbb{E}[\Tcmp_i]/\mathbb{E}[\Tu_m(n)]$ denote the ratio of the expected computation latency $\mathbb{E}[\Tcmp_i]$ of an arbitrary device $i\in[n]$ (where workers have the same latency distribution as in Section \ref{latency_formulation}) to the expected latency $\mathbb{E}[\Tu_m(n)]$.
\begin{proposition}[Device Failure Scenario]\label{prop:fail}

Under the conditions in Proposition \ref{prp:cp2uncoded}, when $n\geq k+1$, $R^{\textrm{cmp}}(n)>0.1$, and one device failure occurs, there exists a $k^*_{sub}\in (1, n-1)$ such that 
$\mathbb{E}[\Tu_m(n)]-\mathbb{E}[\Tc_m(n,k^*_{sub})]>\Delta$. 
\end{proposition}

 Based on Propositions \ref{prp:cp2uncoded} and \ref{prop:fail}, as the increase of straggling/failure, 
 $\mathbb{E}[\Tc(n)]<\mathbb{E}[\Tu(n)]$ always holds, and CoCoI offers more latency reduction compared to the uncoded.

\begin{figure}
    \centering
\includegraphics[height=3.2cm]{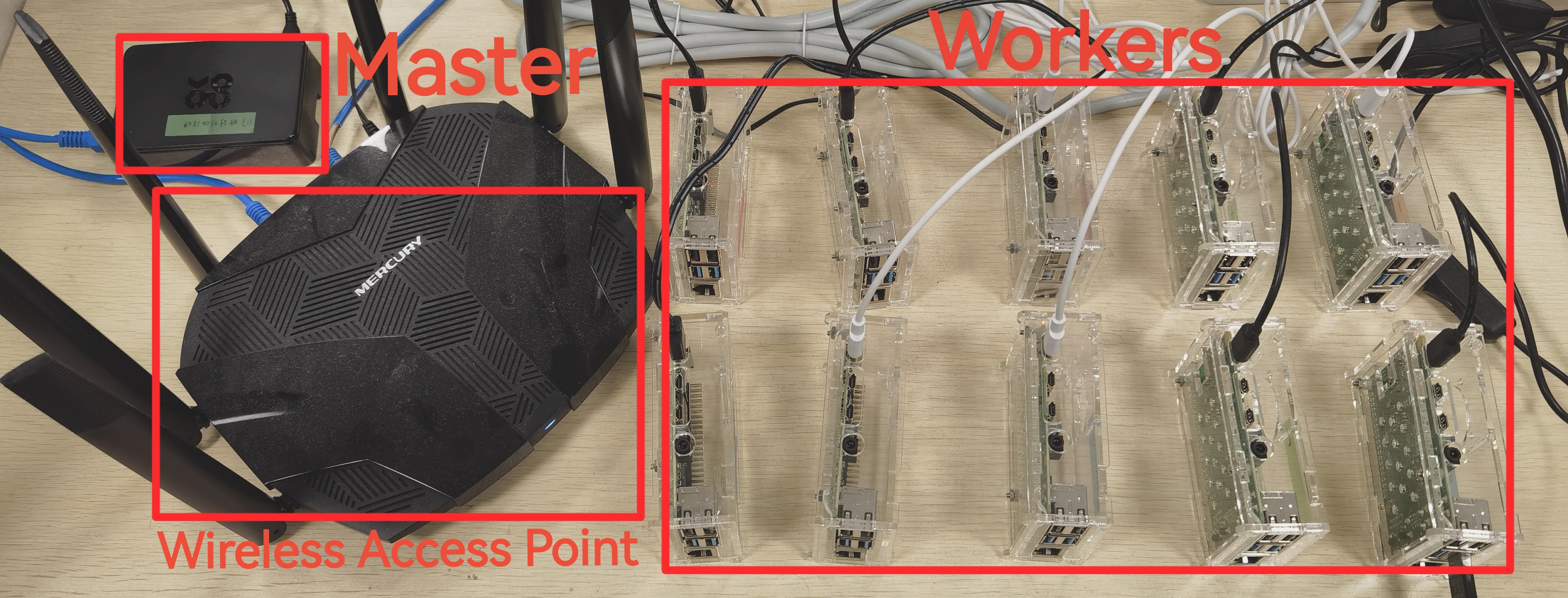}\vspace{-1mm}
    \caption{Testbed for CoCoI system.\vspace{-5mm}}
    \label{fig:testbet}
\end{figure}
\section{Performance Evaluation}\label{sec:experiment}


We build a testbed to implement CoCoI system, as shown in Fig. \ref{fig:testbet}. This testbed consists of a wireless access point Mercury D128 (2.4/5GHz WiFi, 1200 Mbps) to provide WIFI access and 11 Raspberry-Pi 4Bs with Quad Core ARM Cortex-A72 1.5GHz. One of the Raspberry-Pi 4Bs is set as the master device with wired connection, and the remaining $n=10$ devices are workers with WIFI connection. 
The operating system of the master and worker devices is Raspberry Pi OS (64-bit) based on linux and the architecture is AArch64. We apply the widely-used deep learning framework PyTorch-CPU (version 1.11.0) to support CNN inference on Raspberry-Pi. We evaluate well-known CNNs VGG16 and ResNet18 on testbed for inference with $224\times224\times3$ images as input data. Each experiment has been conducted for 20 times.

We consider three scenarios to evaluate the performance of CoCoI. Since
Raspberry-Pi 4Bs originally have similar processing and communication capacities, we manually put devices into sleep in scenario-1 for evaluating the impact of  straggling effect. Scenario-2 and scenario-3 are more practical scenarios without manually introduced sleeping behaviors.
\begin{itemize}
    \item Scenario 1 (Straggling):
    We introduce additional random delay to wireless transmission, which  follows exponential distribution with a scale of $\lambda_{\text{tr}}\overline{T^{\text{tr}}}$. Here,  $\overline{T^{\text{tr}}}$ is the original expected transmission latency.
    \item Scenario 2 (Device Failure): $n_f$ workers randomly fail in each turn of subtask execution. 
    \item Scenario 3 (Straggling and Failure): Based on scenario-2, one worker is a ``high-probability" straggler, which usually has larger execution latency then other workers. 
\end{itemize}
For reference, we measure the latency when CNN inference is performed locally on a single Raspberry Pi 4B. For VGG16, the average latency of normal worker and ``high-probability" straggler are 50.8s and 85.2s, respectively. For ResNet18, the associated latency are  89.8s and 148.8s, respectively.

We consider the following methods for comparison.
In all methods, type-1 tasks are executed in a Map-Reduce style.
\begin{itemize}
    \item CoCoI-$k^*$: CoCoI with the optimal $k^*$, which is obtained by testing all feasible $k$'s and choose the best one.
    \item CoCoI-$k^\circ$:  CoCoI with the approximate optimal $k^{\circ}$.
    \item Uncoded \cite{dina}: Each type-1 task is split into $n$ subtasks and allocated to workers. If any worker fails, the subtask will be re-assigned to another worker for execution.
    \item Replication \cite{ciucu2021Prac}: Each type-1 task is split into $k=\lfloor{n}/{2}\rfloor$ subtasks, with each subtask assigned to $2$ workers for repeated execution. The master determines the final result upon receiving one copy of each subtask's output. 
    \item LtCoI-$k_l$: CoCoI while incorporating a Luby Transform (LT)-based  coding scheme \cite{fang2023latency}. The splitting strategy $k_l=W_\textsc{O}$ represents the finest-grained splitting of type-1 tasks. Note that $k_l$ can be larger than $n$.
    \item LtCoI-$k_s$: CoCoI while incorporating an  LT-based  coding scheme \cite{fang2023latency}, with the optimal splitting strategy $k_s$ under $k_s\leq n$. This is introduced to better compare with MDS codes in a similar number of source tasks $k$.
    
\end{itemize}
Note that LT codes (see Appendix \ref{LtCoI_detail})
is a type of rateless fountain codes\cite{fang2023latency}, which is the recent state-of-the-art coding scheme.
Comparing our approach with LtCoI-$k_l$ and LtCoI-$k_s$ shows the rationale of using MDS coding scheme.
\begin{figure}[t]
    \centering
        \includegraphics[width=8.7cm]{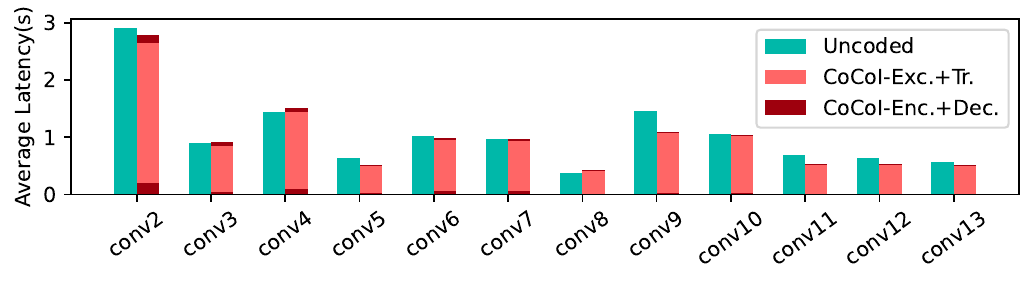}\vspace{-4mm}\\
        (a)\\
        \includegraphics[width=8.7cm]{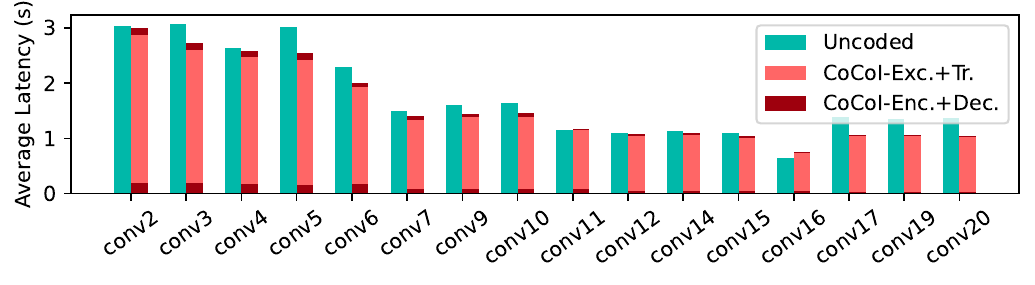}\vspace{-3mm}\\
        (b)\vspace{-3mm}
    \label{fig:CNN_latency_lambda_vgg16}
    \caption{Inference latency of convolutional  layers of (a) VGG16 and (b) ResNet18 under secnario-1 with $\lambda^{\text{tr}}=0.5$.}\label{fig:encod}
\end{figure}

\newcolumntype{C}[1]{>{\centering\arraybackslash}p{#1}}
\begin{table}[t]
\caption{Statistic of $k^*$ and $k^\circ$ under Scenario-1}
\label{tab:optimal_estimated}
\centering
\begin{tabular}{C{1cm}|C{10.8em}|*{5}{C{0.3cm}}}
\toprule
CNN & Statistics $\backslash$ {$\lambda^{\text{tr}}$} & 0.2 & 0.4 & 0.6 & 0.8 & 1.0 \\
\midrule
\multirow{3}{*}{VGG16} 
& $\max_{l\in \mathcal{L}_d}|k^*_l-k^\circ_l|$  & 1 & 1 & 1 & 1 & 1 \\ 
& $\sum_{l\in \mathcal{L}_d}|k^*_l-k^\circ_l|/n_l$ & 0.42 & 0.5 & 0.5 & 0.42 & 0.5 \\ 
& $\sum_{l\in \mathcal{L}_d}|t^*_l-t^\circ_l|$\ (in\ sec.) & 0.10 & 0.08 & 0.11 & 0.49 & 0.51 \\
\midrule
\multirow{3}{*}{ResNet18} 
& $\max_{l\in \mathcal{L}_d}|k^*_l-k^\circ_l|$ & 1 & 1 & 2 & 1 & 1 \\ 
& $\sum_{l\in \mathcal{L}_d}|k^*_l-k^\circ_l|/n_l$ & 0.19 & 0.31 & 0.37 & 0.37 & 0.6 \\ 
& $\sum_{l\in \mathcal{L}_d}|t^*_l-t^\circ_l|$\ (in\ sec.) & 0.35 & 0.15 & 0.97 & 0.37 & 1.3 \\
\bottomrule
\end{tabular}
\end{table}

\begin{figure}[t]
    \centering
     \includegraphics[width=4.15cm]{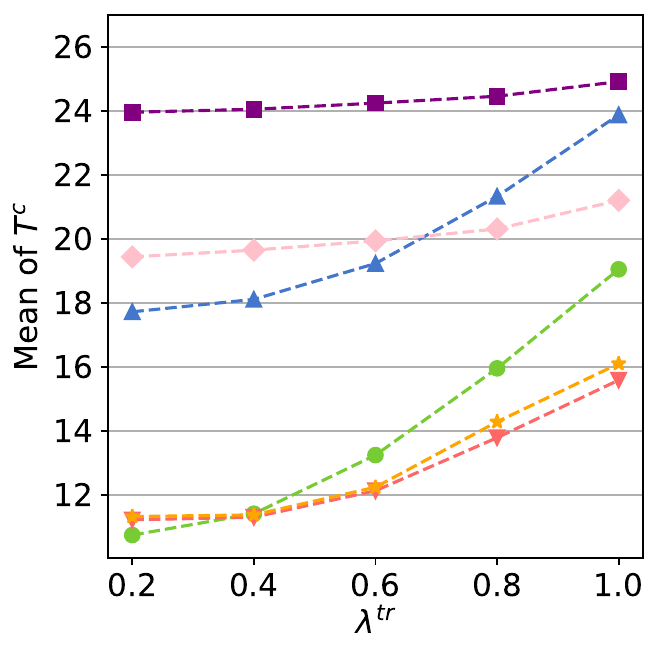}
      \includegraphics[width=4.15cm]{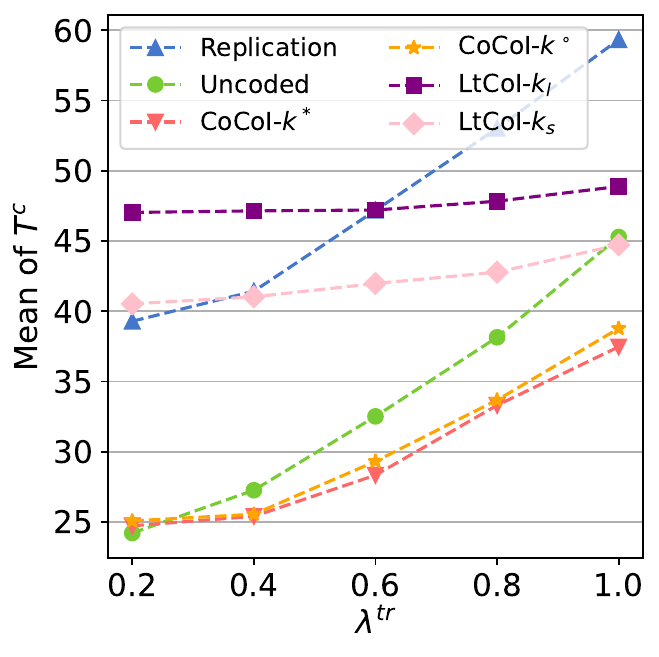}\vspace{-3mm}\\
      \qquad(a)\qquad\ \ \qquad\qquad\qquad\qquad(b)\vspace{-3mm}
    \label{fig:CNN_latency_lambda_resnet}
    \caption{CNN inference latency under scenario-1: (a) VGG16; (b) ResNet18.}
    \label{fig:cnn_inference_lambda}\vspace{-2mm}
\end{figure}

\subsection{Encoding and Decoding Overhead}
Fig. \ref{fig:encod} compares the total latency of each convolutional layer in CNN models between CoCoI and uncoded approach. The dark and light red areas indicate the encoding/decoding latency at the master and the execution and input/output transmission latency at workers in CoCoI, respectively. In Fig. \ref{fig:encod}, the encoding and decoding latency occupies around $2\%-9\%$ of the total latency in each layer. In addition, despite such encoding/decoding overhead, our proposed CoCoI can usually achieve a lower average latency than the uncoded approach. 



\subsection{Approximate Optimal Splitting Strategy}

Table \ref{tab:optimal_estimated} compares the difference between optimal 
and our proposed approximate optimal splitting strategies. Specifically, $\mathcal{L}_d$ denotes the set of type-1 tasks in CNN models, where the set has a size of $n_l$. For each layer $l\in\mathcal{L}_d$, let  $k^*_l$ and $k^\circ_l$ denote the optimal and approximate optimal splitting strategies, respectively. Let $t^*_l$ and $t^\circ_l$ denote the associated latency under $k^*_l$ and $k^\circ_l$, respectively. As shown in Table \ref{tab:optimal_estimated}, the maximum difference $|k_l^*-k^\circ_l|$ is usually not larger than one, and the average difference is around 0.5. Meanwhile, the average latency difference is no larger than 1.3 seconds. 

Fig.  \ref{fig:cnn_inference_lambda} compares the CNN inference latency. In particular,  CoCoI-$k^*$ and CoCoI-$k^\circ$ lead to similar inference latency, which validates the rationale of using $k^\circ$ in practical systems. 
\vspace{-2mm}
\subsection{Method Comparison}

\begin{figure}[t]
    \centering
    \includegraphics[width=4.15cm]{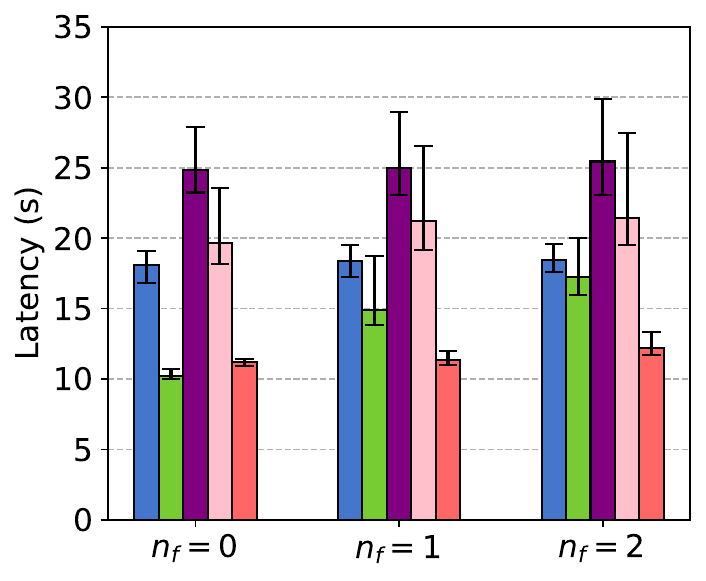}
    \includegraphics[width=4.15cm]{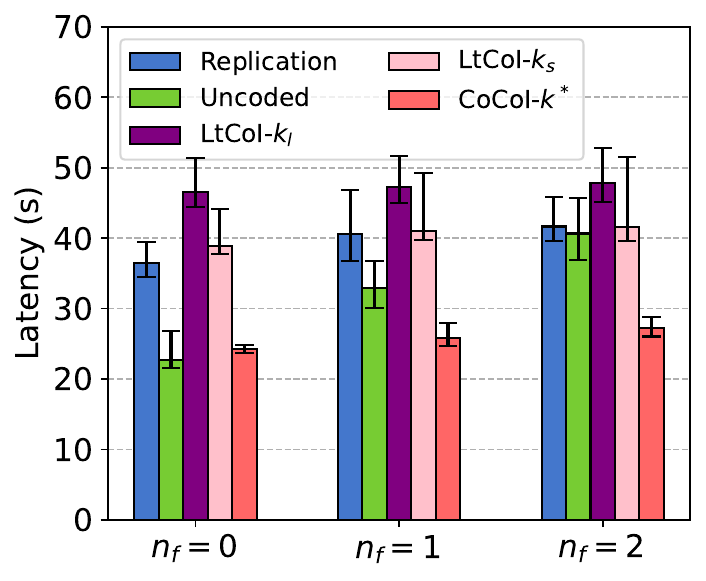}\vspace{-2mm}\\
    \qquad(a)\qquad\qquad\qquad\qquad\qquad\ (b)\\
    \includegraphics[width=4.15cm]{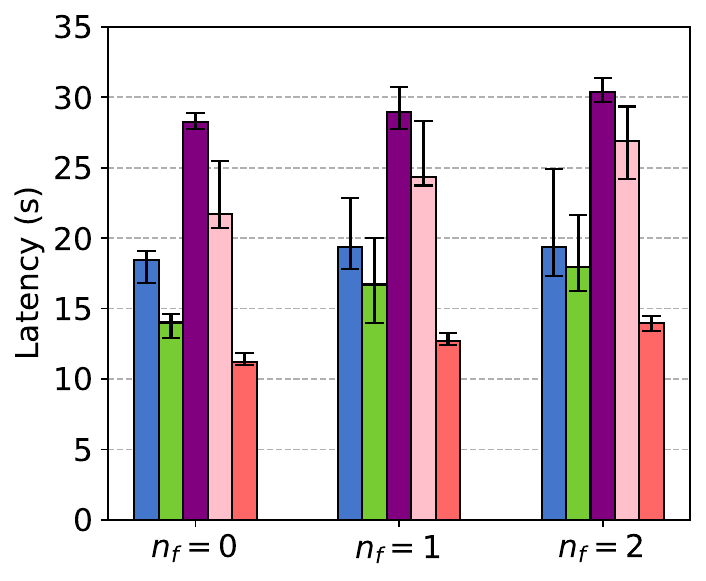}
    \includegraphics[width=4.15cm]{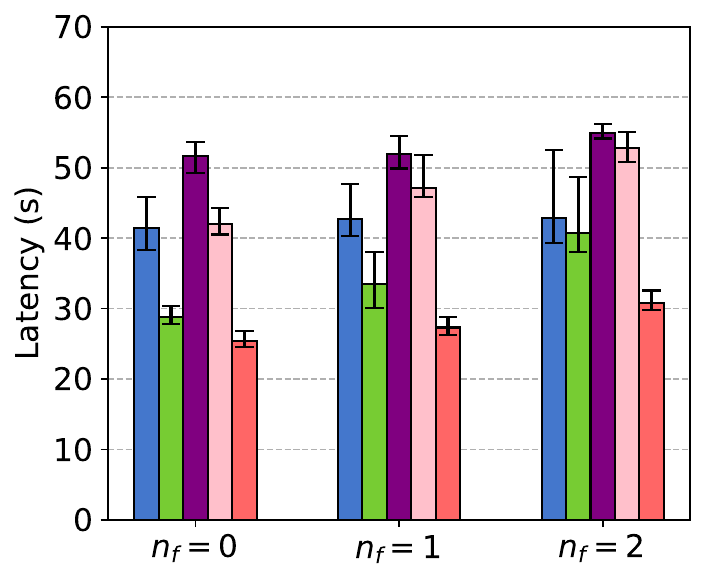}\vspace{-2mm}\\
    \qquad(c)\qquad\qquad\qquad\qquad\qquad\ (d)
    \vspace{-2mm}
    \caption{Inference of (a) VGG16 and  (b) ResNet18 in scenario-2;  inference of (c) VGG16 and (d) ResNet18  in scenario-3.
    }\label{fig:cnn_inference_failure}\vspace{-2mm}
\end{figure}



Fig. \ref{fig:cnn_inference_lambda} shows the CNN inference latency in scenario-1. When $\lambda^{\text{tr}}$ is small (e.g., $\lambda^{\textrm{tr}}\leq 0.2$), ``uncoded" is slightly faster than CoCoI due to the smaller workload at each worker. However, under a moderate degree of straggling effect (e.g., $\lambda_{\text{tr}}\geq0.4$),\footnote{Even under the most severe straggling case (i.e., $\lambda^{\text{tr}}=1$) considered in scenario-1, it corresponds to a moderate degree of straggling effect  in practical systems (less severe than that considered in \cite{coded_convolution, Zhou2022Dynamic}). This case  corresponds to the case with $R>1.3$ in Section \ref{cp2uncoded}. It validates that in addition to our theoretical results on severe straggling scenarios with $R\leq 1$, our approach can reduce the latency under less severe scenarios with $R$ larger than one. 
} both CoCoI-$k^*$ and CoCoI-$k^\circ$ can reduce the inference latency, comparing to the benchmarks. The latency reduction can be up to 20.2\% when $\lambda^{\text{tr}}=1$.
Although LT codes may be more robust under stragglers\cite{fang2023latency}, it achieves worse performance than ``uncoded" and CoCoI in scenario-1.




Fig. \ref{fig:cnn_inference_failure} shows the CNN inference latency in scenario-2 and scenario-3, both of which consider device failure. The error bars here imply the variance of inference latency.
When there is no device failure or straggler (i.e., $n_f=0$ in scenario-2), CoCoI is slightly slower than ``uncoded" due to the system redundancy.
As $n_f$ increases from 0 to 2 in scenario-2, the latency of ``uncoded" increase by 68.3\%-79.2\%, while CoCoI and LtCoI show better resistance to failure. 
However, the fine-grained splitting of LtCoI-$k_l$ introduces excessive transmission overhead, and LtCoI-$k_s$ incurs a higher subtask computation overhead due to the inevitable higher redundancy resulting from the small $k$ in LT codes.
Above all, CoCoI achieves lower and more stable inference latency with less variance (according to the length of error bars) under the existence of device failure. When compared with ``uncoded", the latency reduction can be up to 34.2\% in scenario-2 and 26.5\% scenario-3.

\section{Conclusion}\label{sec:conclusion}
In this work, we proposed a novel distributed coded inference system, called CoCoI. This system  mitigates the straggling and device failure issue by splitting 2D convolution layers, considering the data dependency between high-dimensional input and output data, and use MDS codes to efficiently generate redundancy in distributed inference.
To understand the optimal splitting strategy, we formulated an expected inference latency minimization problem. Despite its non-convexity and lack of an explicit expression, we determined an approximate solution, whose empirical performance is close to the optimal solution. We proved that when compared with uncoded method, CoCoI can reduce the inference latency under a certain degree of straggling and device failure. 
Evaluations on the real-world testbed shows that CoCoI can effectively reduce inference latency in the presence of stragglers and device failure.
An interesting future direction is to optimize the subtask allocation across heterogeneous workers for further  inference latency reduction.

\bibliographystyle{IEEEtran}
\bibliography{citation}

\begin{thebibliography}{10}
\providecommand{\url}[1]{#1}
\csname url@samestyle\endcsname
\providecommand{\newblock}{\relax}
\providecommand{\bibinfo}[2]{#2}
\providecommand{\BIBentrySTDinterwordspacing}{\spaceskip=0pt\relax}
\providecommand{\BIBentryALTinterwordstretchfactor}{4}
\providecommand{\BIBentryALTinterwordspacing}{\spaceskip=\fontdimen2\font plus
\BIBentryALTinterwordstretchfactor\fontdimen3\font minus
  \fontdimen4\font\relax}
\providecommand{\BIBforeignlanguage}[2]{{%
\expandafter\ifx\csname l@#1\endcsname\relax
\typeout{** WARNING: IEEEtran.bst: No hyphenation pattern has been}%
\typeout{** loaded for the language `#1'. Using the pattern for}%
\typeout{** the default language instead.}%
\else
\language=\csname l@#1\endcsname
\fi
#2}}
\providecommand{\BIBdecl}{\relax}
\BIBdecl

\bibitem{zhang2023computer}
Y.-J. Zhang, ``Computer vision overview,'' in \emph{3-D Computer Vision:
  Principles, Algorithms and Applications}.\hskip 1em plus 0.5em minus
  0.4em\relax Springer, 2023, pp. 1--35.

\bibitem{Shuvo2023EfficientAO}
M.~M.~H. Shuvo, S.~K. Islam, J.~Cheng, and B.~I. Morshed, ``Efficient
  acceleration of deep learning inference on resource-constrained edge devices:
  A review,'' \emph{Proceedings of the IEEE}, vol. 111, pp. 42--91, 2023.

\bibitem{vgg}
K.~Simonyan and A.~Zisserman, ``Very deep convolutional networks for
  large-scale image recognition,'' in \emph{3rd International Conference on
  Learning Representations, {ICLR} 2015}, Y.~Bengio and Y.~LeCun, Eds., 2015.

\bibitem{resnet}
K.~He, X.~Zhang, S.~Ren, and J.~Sun, ``Deep residual learning for image
  recognition,'' in \emph{Proceedings of IEEE Conference on Computer Vision and
  Pattern Recognition (CVPR)}, 2016, pp. 770--778.

\bibitem{Hu2019DADS}
C.~Hu, W.~Bao, D.~Wang, and F.~Liu, ``Dynamic adaptive {DNN} surgery for
  inference acceleration on the edge,'' in \emph{IEEE INFOCOM 2019 - IEEE
  Conference on Computer Communications}, 2019, pp. 1423--1431.

\bibitem{yang2023ondemand}
L.~Yang, X.~Shen, C.~Zhong, and Y.~Liao, ``On-demand inference acceleration for
  directed acyclic graph neural networks over edge-cloud collaboration,''
  \emph{Journal of Parallel and Distributed Computing}, vol. 171, no.~C, p.
  79–87, Jan. 2023.

\bibitem{modnn}
J.~Mao, X.~Chen, K.~W. Nixon, C.~Krieger, and Y.~Chen, ``{MoDNN}: Local
  distributed mobile computing system for deep neural network,'' in
  \emph{Proceedings of Design, Automation \& Test in Europe Conference \&
  Exhibition (DATE), 2017}, 2017, pp. 1396--1401.

\bibitem{dina}
T.~Mohammed, C.~Joe-Wong, R.~Babbar, and M.~D. Francesco, ``Distributed
  inference acceleration with adaptive {DNN} partitioning and offloading,'' in
  \emph{Proceedings of IEEE Conference on Computer Communications (INFOCOM)},
  2020, pp. 854--863.

\bibitem{coedge}
L.~Zeng, X.~Chen, Z.~Zhou, L.~Yang, and J.~Zhang, ``{CoEdge}: Cooperative {DNN}
  inference with adaptive workload partitioning over heterogeneous edge
  devices,'' \emph{IEEE/ACM Transactions on Networking}, vol.~29, no.~2, pp.
  595--608, 2021.

\bibitem{bitar2020stochastic}
R.~Bitar, M.~Wootters, and S.~El~Rouayheb, ``Stochastic gradient coding for
  straggler mitigation in distributed learning,'' \emph{IEEE Journal on
  Selected Areas in Information Theory}, vol.~1, no.~1, pp. 277--291, 2020.

\bibitem{zhu2023heterogeneous}
H.~Zhu, L.~Chen, X.~Chen, and W.~Wang, ``Heterogeneous secure coded matrix
  multiplication: Straggler problem versus information leakage,'' in \emph{2023
  IEEE 98th Vehicular Technology Conference (VTC2023-Fall)}, 2023, pp. 1--6.

\bibitem{said2022optimized}
S.~A. Said, S.~M. Habashy, S.~A. Salem, and E.~M. Saad, ``An optimized
  straggler mitigation framework for large-scale distributed computing
  systems,'' \emph{IEEE Access}, vol.~10, pp. 97\,075--97\,088, 2022.

\bibitem{tran2023disco}
H.~Tran-Dang and D.-S. Kim, ``Disco: Distributed computation offloading
  framework for fog computing networks,'' \emph{Journal of Communications and
  Networks}, vol.~25, no.~1, pp. 121--131, 2023.

\bibitem{behrouzi2019data}
A.~Behrouzi-Far and E.~Soljanin, ``Data replication for reducing computing time
  in distributed systems with stragglers,'' in \emph{2019 IEEE International
  Conference on Big Data (Big Data)}, 2019, pp. 5986--5988.

\bibitem{ciucu2021Prac}
F.~Ciucu, F.~Poloczek, L.~Y. Chen, and M.~Chan, ``Practical analysis of
  replication-based systems,'' in \emph{IEEE INFOCOM 2021 - IEEE Conference on
  Computer Communications}, 2021, pp. 1--10.

\bibitem{Lee2018speeding}
K.~Lee, M.~Lam, R.~Pedarsani, D.~Papailiopoulos, and K.~Ramchandran, ``Speeding
  up distributed machine learning using codes,'' \emph{IEEE Transactions on
  Information Theory}, vol.~64, no.~3, pp. 1514--1529, 2018.

\bibitem{Mallick2022rateless}
A.~Mallick, M.~Chaudhari, U.~Sheth, G.~Palanikumar, and G.~Joshi, ``Rateless
  codes for near-perfect load balancing in distributed matrix-vector
  multiplication,'' \emph{Communications of the ACM}, vol.~65, no.~5, p.
  111–118, Apr. 2022.

\bibitem{coded_convolution}
S.~Dutta, V.~Cadambe, and P.~Grover, ``Coded convolution for parallel and
  distributed computing within a deadline,'' in \emph{2017 IEEE International
  Symposium on Information Theory (ISIT)}, 2017, pp. 2403--2407.

\bibitem{Zhou2022Dynamic}
B.~Zhou, J.~Xie, and B.~Wang, ``Dynamic coded distributed convolution for
  {UAV}-based networked airborne computing,'' in \emph{2022 International
  Conference on Unmanned Aircraft Systems (ICUAS)}, 2022, pp. 955--961.

\bibitem{fang2023latency}
B.~Fang, K.~Han, Z.~Wang, and L.~Chen, ``Latency optimization for {Luby
  Transform} coded computation in wireless networks,'' \emph{IEEE Wireless
  Communications Letters}, vol.~12, no.~2, pp. 197--201, 2023.

\bibitem{pytorch_docs}
{PyTorch}, ``{PyTorch Documentation},'' 2019, [Online]. Available:
  \url{https://pytorch.org/docs/1.11/index.html}.

\bibitem{marshall1967multivariate}
A.~W. Marshall and I.~Olkin, ``A multivariate exponential distribution,''
  \emph{Journal of the American Statistical Association}, vol.~62, no. 317, pp.
  30--44, 1967.

\bibitem{hcmm}
S.~Kianidehkordi, N.~Ferdinand, and S.~C. Draper, ``Hierarchical coded matrix
  multiplication,'' \emph{IEEE Transactions on Information Theory}, vol.~67,
  no.~2, pp. 726--754, 2021.

\bibitem{sharedcoded}
Y.~Sun, F.~Zhang, J.~Zhao, S.~Zhou, Z.~Niu, and D.~Gündüz, ``Coded
  computation across shared heterogeneous workers with communication delay,''
  \emph{IEEE Transactions on Signal Processing}, vol.~70, pp. 3371--3385, 2022.

\bibitem{david2004order}
H.~A. David and H.~N. Nagaraja, \emph{Order statistics}.\hskip 1em plus 0.5em
  minus 0.4em\relax John Wiley \& Sons, 2004.

\end{thebibliography}

\clearpage
\appendix

\subsection{Bottleneck of CNN Inference}\label{app:bottleneck}

\begin{figure}[t]
\centering
\includegraphics[width=0.48\textwidth]{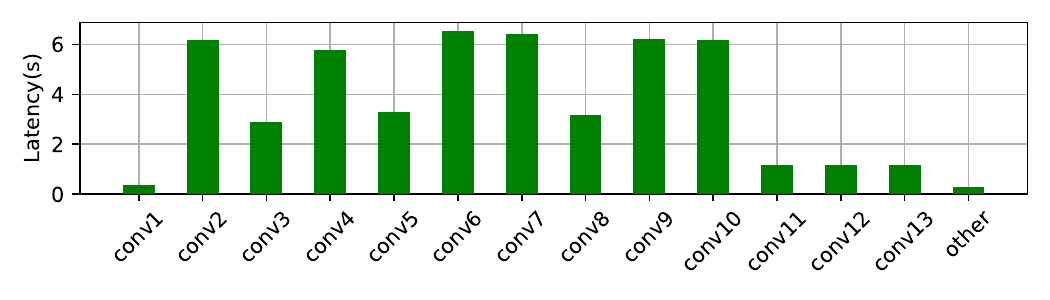}\vspace{-0.5mm}\\
(a) VGG16\\  
    \includegraphics[width=0.48\textwidth]{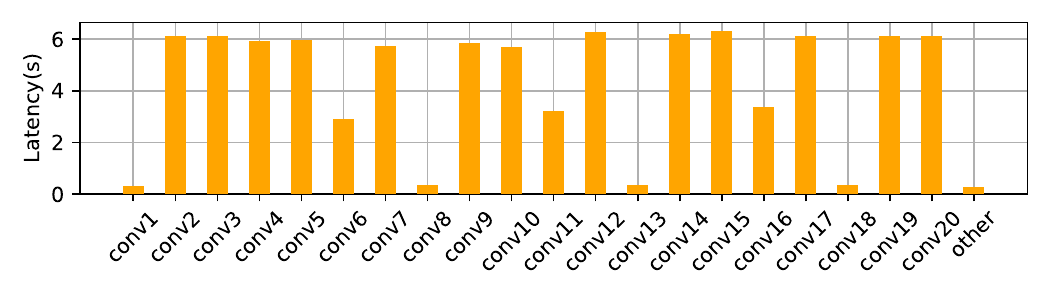}\vspace{-1mm}\\
   (b) ResNet18
\vspace{-2mm}
\caption{Inference latency at different layers. The total latency on VGG16 and ResNet18 are 50.8s and 89.8s, respectively.\vspace{-2mm}}\label{fig:expample}
\label{fig:cnn_layer_latency}
\end{figure}

Figure \ref{fig:expample} shows the local inference latency of VGG16\cite{vgg} and ResNet18\cite{resnet} \textbf{by layer} on a single Raspberry Pi 4B under the same software settings as that of Section \ref{sec:experiment}. Here, ``other" denotes the inference latency of all layers except for convolutional layers, including pooling, activation, normalization, and linear layers. It takes 50.8s for a complete local inference of VGG16, and 89.8s for ResNet18. While convolutional layers account for 99.43\% and 99.68\% of total inference latency, respectively, demonstrating the evidently high computational complexity of convolutional layers. Note that \textbf{not all} convolutional layers are type-1 layers, for example, ``conv1" in VGG16 and ``conv1", ``conv8", ``conv13", ``conv18" in ResNet18.
We classify a layer to be a type-1 layer according to whether performing distributed execution on that layer \textbf{can} accelerate its completion latency.

\subsection{Stochastic Latency and Exponential Distribution Fitting}\label{exponential_fitting}
To justify the modeling of the wireless transmission or computation latency using a shift-exponential distribution, we evaluate the wireless device-to-device transmission latency with a bandwidth limitation of 100Mbps and the convolution execution latency on workers on our testbed (with Raspberry Pi 4B using python 3.9 and cpu version of torch-1.11).

To  evaluate the wireless transmission latency, the master sends a 2MB torch tensor to workers for  500 times, and the workers response with a short message immediately for each received data. The master records the RTT until receiving the corresponding response. 
To evaluate the execution latency, we let all 10 workers repeatedly execute the same 2D-convolution task (specifically, the third convolutional layer of VGG16) for 100 times, recording each computation delay.

\begin{figure}[t]
    \centering
    \includegraphics[width=0.48\linewidth]{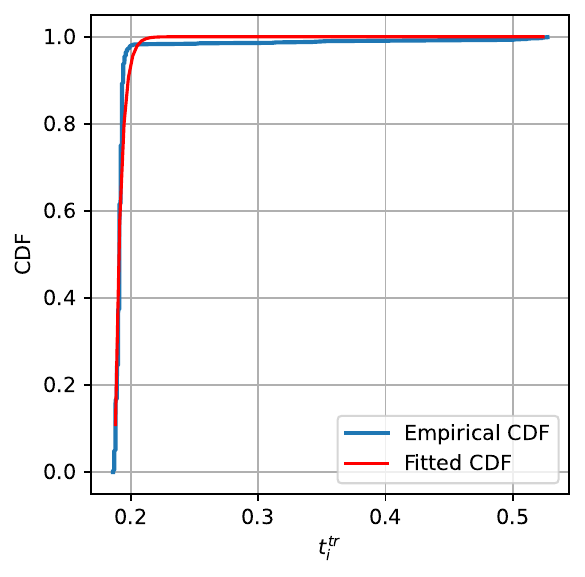}
    \label{fig:comm_latency_fit}
    \includegraphics[width=0.48\linewidth]{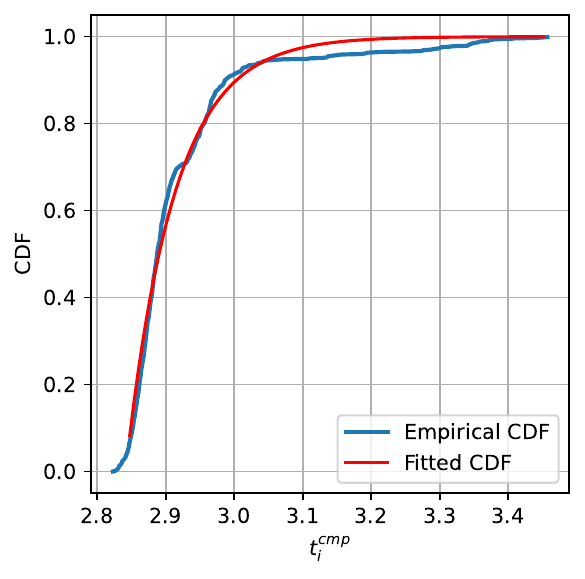}
    \label{fig:comp_latency_fit}
    \vspace{-2mm}\\  
    \qquad(a)\qquad\ \ \ \qquad\qquad\qquad\qquad(b)\ \ \vspace{-2mm}
    \caption{CDF of Empirical (a) transmission and (b) computation latency.}
    \label{fig:latency_fit}
\end{figure}

Figure \ref{fig:latency_fit} shows the CDF curves of the collected transmission and computation latency. We use the shift-exponential distribution to fit these CDF curves. As shown in Figure \ref{fig:latency_fit}, the empirically measured transmission and computation latency distributions show a high degree of consistency with the shift-exponential distribution. Thus, we use shift-exponential distribution to model the latency in CoCoI.

\begin{table}[t]
\caption{Notations for Key Parameters}
\vspace{-2mm}
\centering
\begin{tabularx}{\linewidth}{l|l}
Notation & Explanation\\
\midrule
$n,k$ & Number of worker devices/source subtasks \\
$f(\cdot)$ & 2D convolution with convolution kernel\\
$\mathbf{I},\mathbf{O}$ & Input/Output feature map of convolution, $\mathbf{O}=f(\mathbf{I})$\\
$C_\textsc{I}, C_\textsc{O}$ & \textit{In/Out\_channels} of convolution kernel\\
$K_W,S_W$ & \textit{Kernel\_size/Stride} values (on the width dimension)\\
$W_\textsc{I}, W_\textsc{O}$ & Width of $\mathbf{I},\mathbf{O}$\\
$W^p_\textsc{I}, W^p_\textsc{O}$ & Width of input/output partition\\
$\mum,\thetam$ & Straggling and shift coefficients of comp. on the master\\
$\mucmp,\thetacmp$ & Straggling and shift coefficients of comp. on workers\\
$\murec,\thetarec$ & Straggling and shift coefficients of recv. on workers\\
$\musen,\thetasen$ & Straggling and shift coefficients of send. on workers\\
$\Trec_i,\Tcmp_i,\Tsen_i$ & Recv./Comp./Send. latency of subtask $i\in[n]$\\
$\Tw_{n:k}$ & Completion latency of the $k^{\text{th}}$ fastest subtask among $n$\\
$\Tc$ & Overall latency of a distributed execution\\
$k^*, k^\circ$ & Optimal $k$ to achieve lowest $\mathbb{E}[\Tc]$ and $L(k)$\\
\bottomrule
\end{tabularx}
\label{tab:params}
\end{table}

\subsection{Proof for Lemma \ref{lem:convex}}\label{proof_convex}
The objective function in \eqref{eq:p2} can be represented by  function $P(k)$ plus a constant value: 
\begin{multline}
    P(k)\triangleq h_1(\mum, \thetam)k+h_2(\thetarec, \thetasen,\thetacmp)\frac{1}{k}\\
    +h_3(\murec,\musen,\mucmp)\frac{1}{k}\ln\frac{n}{n-k}+h_4(\murec)\ln\frac{n}{n-k}.
\end{multline}
Recall  $h_1(\mum,\thetam)\!\triangleq\!2(\frac{1}{\mum}\!+\!\thetam)(nI_{ov}\!+\!O)$, $h_2(\thetarec,\thetasen,\thetacmp)\!\triangleq\!4I_W\thetarec\!+\!4O\thetasen+N_{c}\thetacmp$, $h_3(\murec,\musen,\mucmp)\triangleq\ \frac{4I_W}{\murec}+\frac{4O}{\musen}+\frac{\Ncmp_{t}}{\mucmp}$, $ h_4(\murec)\triangleq\ \frac{4I_{ov}}{\murec}$, 
with $I_{ov}\triangleq C_\textsc{I}H_\textsc{I}(K-S)$, $I_W\triangleq C_\textsc{I}H_\textsc{I}W_\textsc{O}S$, $O\triangleq C_\textsc{O}H_\textsc{O}W_\textsc{O}$, and $\Ncmp_{t}\triangleq 2C_\textsc{O}H_\textsc{O}C_iK^2W_\textsc{O}$.

Now, we are ready to prove Lemma \ref{lem:convex}. 
In $P(k)$, the terms of $k$, ${1}/{k}$, and  $\ln({n}/{(n-k)})$  are obviously convex by checking their second-order derivatives in $k\in[1,n)$. 
In this proof, we aim to prove that $f(k)\triangleq \ln({n}/{(n-k)})/k$ is also convex for $k\in[1,n)$, 
under which problem \eqref{eq:p2} is convex. 

To prove the convexity of $f(k)$, we need to show that the second-order derivative $f^{''}(k)={(3k-2n)}/{(k^2(n-k)^2)}+{2\ln({n}/{(n-k)})}/{k^3}>0$ for $k\in[1,n)$. Since $k\geq1$, $n>k$, and $n\geq3$, it is equivalent to prove $g(k,n)\triangleq 3k^2-2nk+2\ln({n}/{(n-k)})(n-k)^2>0$. To prove $g(k,n)>0$, we need to show that (i) $g(1,n)>0$ and (ii) $g(k,n)$ is increasing in $k\in[1,n)$. (i) When $k=1$, $g(1,n)=3-2n+2\ln\frac{n}{n-1}(n-1)^2$. It is easy to check that $g(1,n_l)>0$ for a large $n_l$. Then, since $\partial g(1,n)/\partial n<0$, $g(1,n)>0$ is proven. 
(ii) We denote $g(k,n)$ as $g(k)$ and show $g(k)$ is increasing in $k\in[1,n)$. We need to prove  $g'(k)=4((k-n){n}/{(n-k)}+k)>0$.  This is equivalent to prove ${n}/{(n-k)}-1>\ln({n}/{(n-k)})$. Let $x\triangleq{n}/{(n-k)},x\in[n/(n-1),\infty)$. Since 
$x-1>\ln x$ for $x\in[n/(n-1),\infty)$, we have $g'(k)>0$.



\subsection{Empirical Evaluation for Approximation}\label{appendix:eval}

Since the original latency formulation \eqref{P1} does not have an explicit expression, we perform large-scale numerical simulations with the scale of $3\times10^5$ to verify its property.

\begin{figure}[t]
    \centering
   \includegraphics[height=3.7cm]{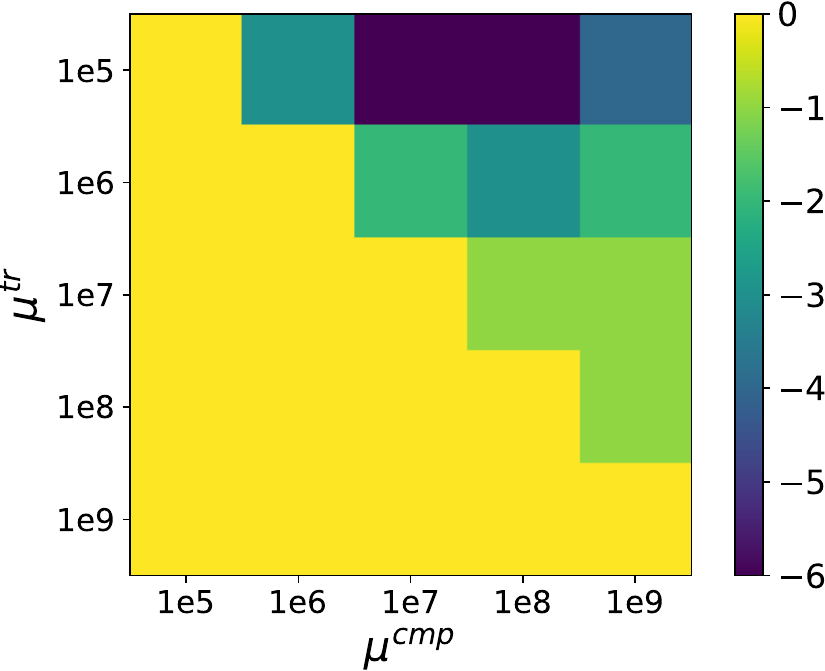}
        \includegraphics[height=3.7cm]{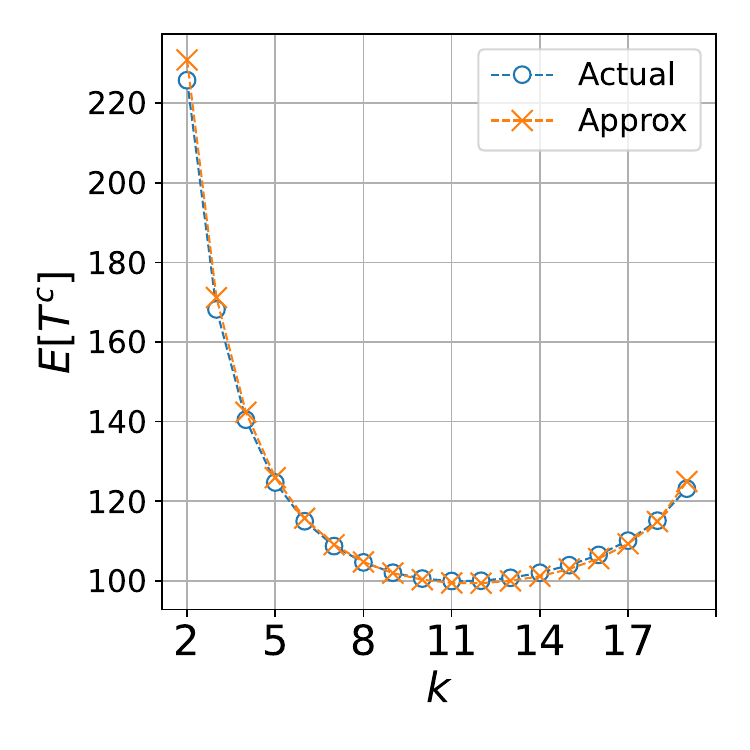}\vspace{-1mm}\\
     \qquad(a) \qquad\qquad\qquad\qquad\qquad\qquad(b)\vspace{-3mm}
    \caption{(a) The difference between the approximate optimal solution $k^{\circ}$ and the optimal splitting strategy $k^*$ to problem \eqref{P1}. We set $\murec=\musen=\mutr$.
    (b) The approximation gap of the latency under different values of $k$ ($\mutr=10^7$ and $\mucmp=10^8$).}
    \label{diff_approx}
    \vspace{-4mm}
\end{figure}

We empirically show the gap between the $k^*$ from \eqref{P1} and $k^\circ$ from \eqref{eq:p2} under different values of $\murec=\musen=\mutr$ and $\mucmp$ in Fig. \ref{diff_approx}.\footnote{Since the $\theta$'s correspond to the shift coefficients, they have minor effect on the approximation error. Thus, we omit the results related to $\theta$'s.}
Let $n=20$.  
In Fig. \ref{diff_approx} (a),  the approximation gap is around zero when the processing capacities of workers have larger variance (i.e., smaller values of $\mucmp$).  
Fig. \ref{diff_approx} (b) shows that the difference between the objective function in \eqref{P1} (denoted by ``Actual") and approximate objective function in \eqref{eq:p2} (denoted by ``Approx") is negligible. A similar observation holds  for all $\mutr$ and $\mucmp$ falling in the yellow range of Fig. \ref{diff_approx} (a).



\begin{figure*}[htbp]
    \centering
        \includegraphics[width=4.3cm]{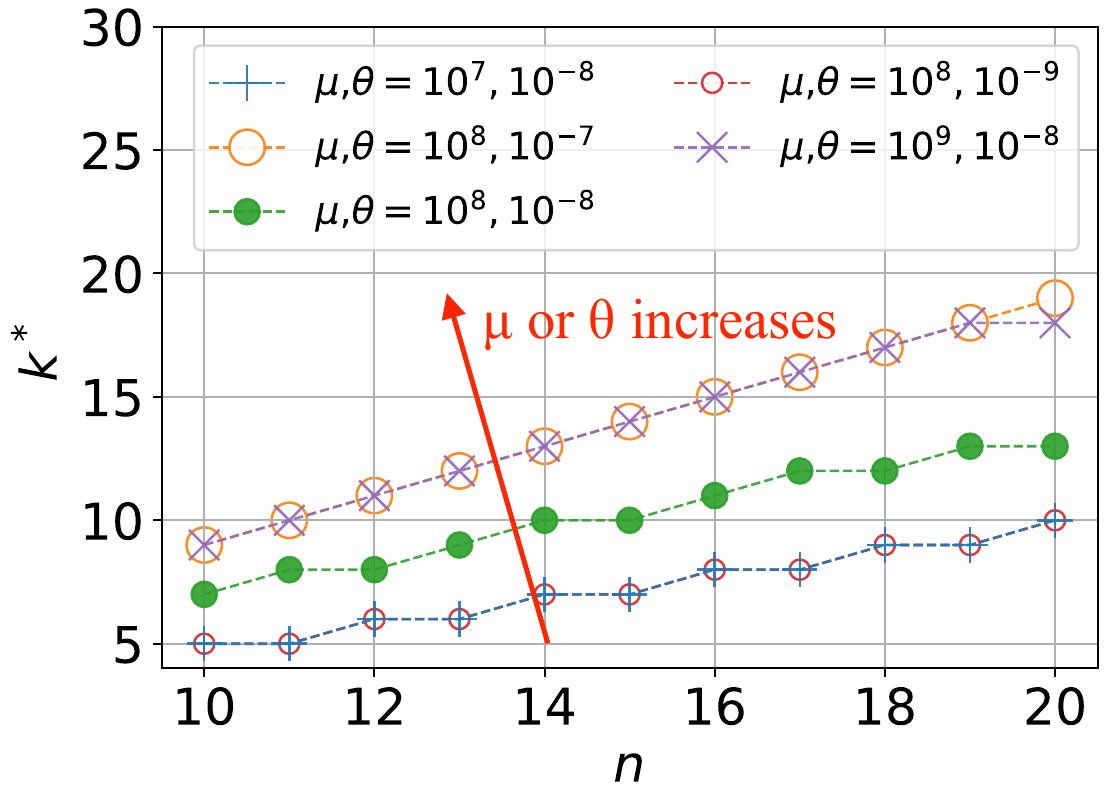}
        \includegraphics[width=4.3cm]{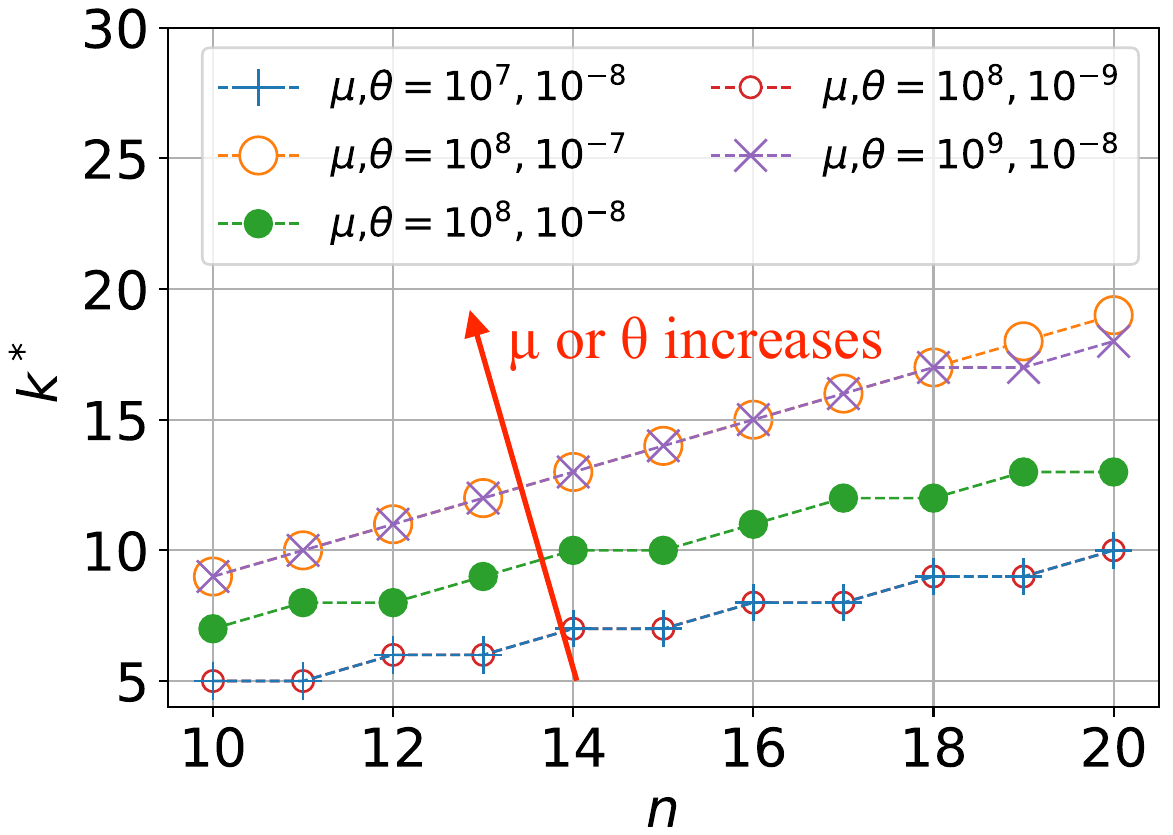}
        \includegraphics[width=4.3cm]{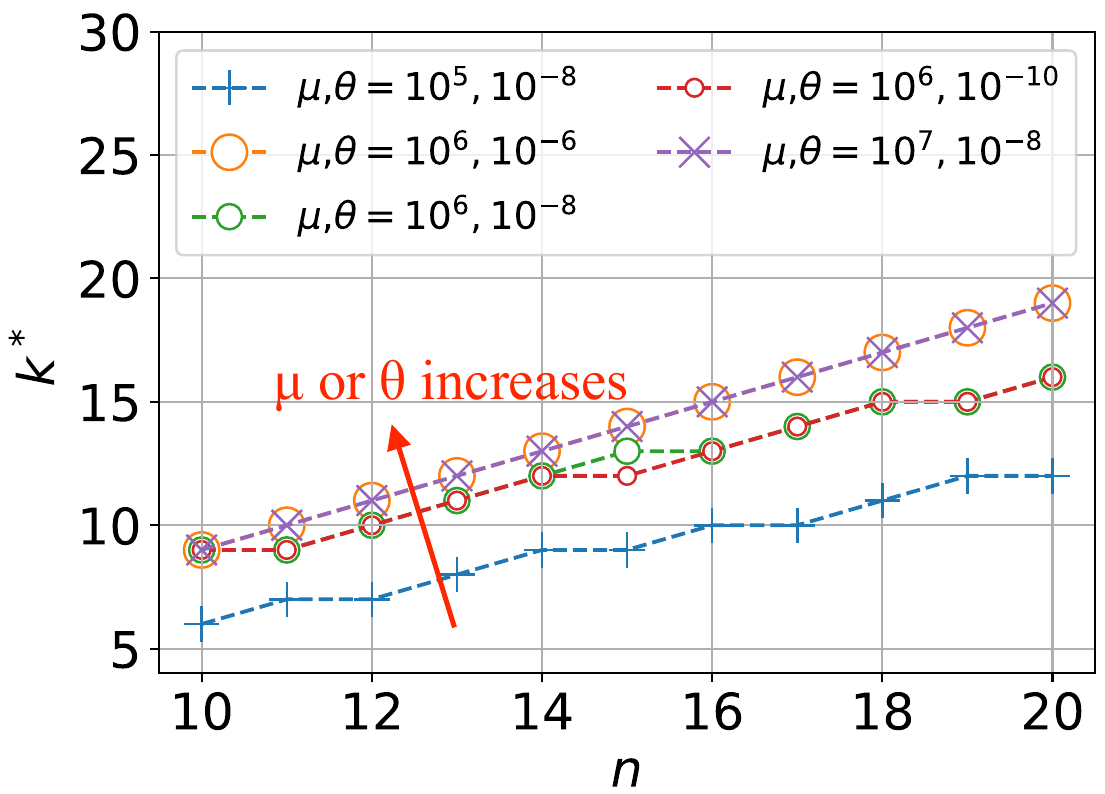}
        \includegraphics[width=4.3cm]{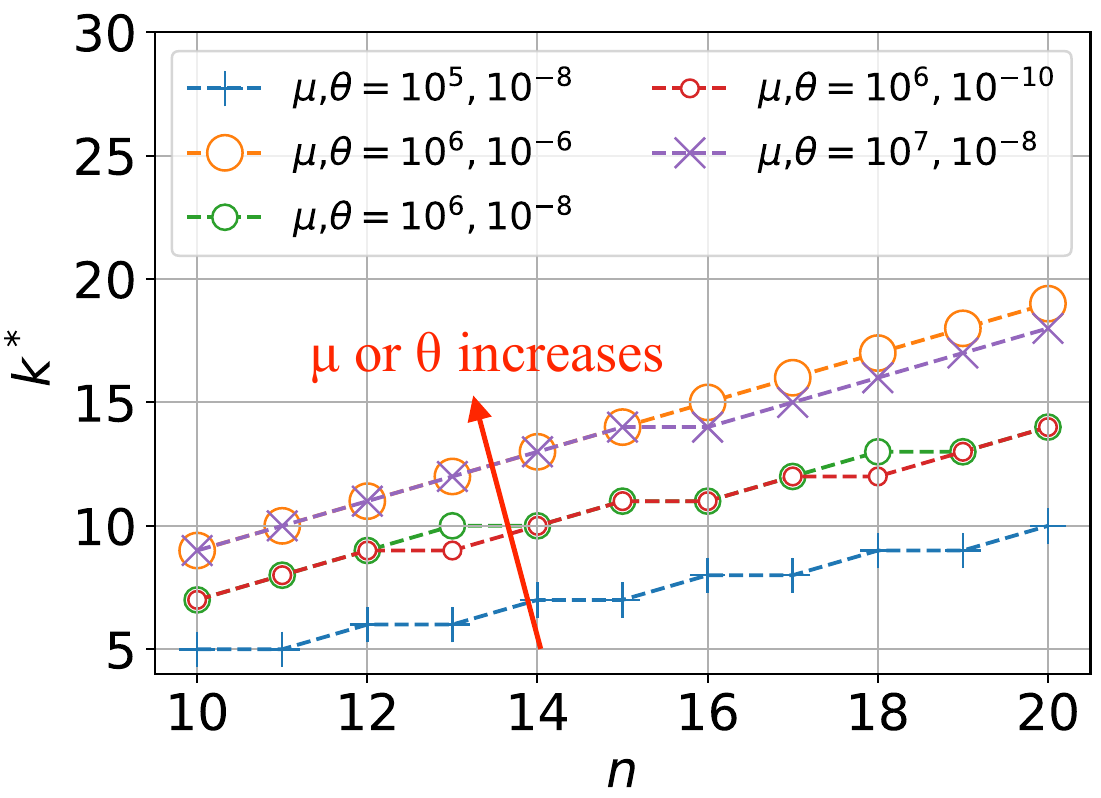}\vspace{-1mm}\\
    \qquad(a) \qquad\qquad\qquad\qquad\qquad\quad\qquad(b)\qquad\qquad\qquad\qquad\qquad\quad\qquad(c)\qquad \qquad\qquad\qquad\qquad\quad\qquad(d)\vspace{-3mm}
    \caption{Impact of $\mu=\mucmp$ and $\theta=\thetacmp$ on (a) actual expected latency and (b) approximate expected latency;  impact of $\mu=\murec=\musen$ and $\theta=\thetarec=\thetasen$ on (c) actual expected latency and (d) approximate expected latency.\vspace{-2mm}}
    \label{fig:impact-para}
\end{figure*}


\subsection{Proof and Analysis for Proposition \ref{prp:impact}}\label{proof_analysis_impact}

\subsubsection{Proof} According to Lemma \ref{lem:convex}, by checking the first-order condition of $L(k)$,  we obtain the following result.

\begin{lemma}[Optimal Solution to Relaxed Problem \eqref{eq:p2}]\label{prp:optimal}
Under the relaxation of $k\in[1,n)$, the optimal solution to the relaxed problem \eqref{eq:p2}, denoted by  $\hat{k}^\circ$, satisfies 
    \begin{multline}\label{eq:kstar}
\!-\!h_1(\mum, \thetam)\!+\!\frac{h_2(\thetarec, \thetasen, \thetacmp)}{(\hat{k}^\circ)^2} \!=\! \frac{h_4(\murec)}{n\!-\! \hat{k}^\circ}+\\h_3(\murec,\musen,\mucmp)\Big(\!-\!\frac{1}{(\hat{k}^\circ)^2}\ln\frac{n}{n\!-\!\hat{k}^\circ}\!+\!\frac{1}{\hat{k}^\circ(n\!-\!\hat{k}^\circ)}\Big).
\end{multline}
\end{lemma}

According to Lemma \ref{prp:optimal}, $\hat{k}^{\circ}$ is the solution that satisfies equality \eqref{eq:kstar}. Let $l(k)$ and $r(k)$ denote the left-hand and right-hand sides of equality \eqref{eq:kstar}, then $\hat{k}^{\circ}$ is the intersection of $l(k)$ and $r(k)$. 
First, we show the intersection is unique. According to lemma \ref{lem:convex}, all the terms in \eqref{eq:tc-approx} are convex regarding $k$, then $-l(k)$ and $r(k)$ are monotonically increasing functions. Besides, $l(k)$ and $r(k)$ have same codomin of $(0,\infty]$ for $k\in [1,n)$, thus their intersection $\hat{k}^{\circ}$ is unique.
Then, the increasing of $\mucmp$ leads to a decreased $h_3(\cdot)$ and hence a smaller
$r(k)$ under each $k\in[1,n)$. Based on the monotonicity of $l(k),r(k)$, their intersection $\hat{k}^{\circ}$ increases accordingly.
In addition, the increasing of  $\thetacmp$ leads to an increased $h_2(\cdot)$ and hence a larger
$l(k)$, which causes the increased $\hat{k}^{\circ}$.
Similar proofs hold for other system parameters and are omitted.

\subsubsection{Insights and Empirical Results} Proposition \ref{prp:impact} provides insights in selecting splitting strategy in practical systems. If any straggler coefficient $\mu$ decreases, then the overall latency has a larger variance and means more severe straggling/failure effect.
Thus, the computation task should be partitioned into fewer pieces (i.e., smaller $k$) to introduce more redundancy (i.e., larger $r\triangleq n-k$). If any shift coefficient $\theta$ of workers (i.e., $\thetacmp, \thetarec,\thetasen$) increases, the minimum completion time of each subtask increases. As a result, the workload of subtasks become larger, so the computation task should be partitioned into smaller pieces (i.e., a larger $k$) to reduce the workload. Finally, if $\frac{1}{\mum}+\thetam$ is larger, the master has a less powerful processing capacity, so $k$ should be decreased to reduce the encoding and decoding latency.

In Figure \ref{fig:impact-para}, we evaluate the impact of system parameters on the optimal splitting strategy for problem \eqref{P1} (see subfigures (a) and (c)) and the approximate optimal strategy $k^*$ determined by problem \eqref{eq:p2} (see subfigures (b) and (d)). First, we observe that the experimental results of the approximate optimal $k^*$ is consistent with the analytical results in Proposition 2. Second, as the total number of workers $n$ increases, the optimal splitting strategy increases under both cases. This is reasonable because a larger $n$ provides a larger worker pool for parallel task execution.

\subsection{Proof for Propositions \ref{prp:cp2uncoded} and \ref{prop:fail}}\label{appendix:performance}
In the following, we first derive the expected latency under uncoded approach. Then, we prove Propositions \ref{prp:cp2uncoded} and \ref{prop:fail}.

\subsubsection{Latency under Uncoded Approach} As in conventional uncoded distributed convolution approach (e.g., \cite{dina}), the master partitions the feature map into $n$ pieces and sends them to workers for execution. The master can obtain the result of each layer only when all $n$ workers send back their output. Based on  \eqref{eq:app} and \cite{david2004order}, the expectation of latency under uncoded approach $\mathbb{E}[\Tu]$ is determined as follows:
\begin{multline}\label{eq:tu-approx}
\mathbb{E}[\Tu(n)]
\approx 
h_2(\thetarec, \thetasen,\thetacmp)\frac{1}{n}+h_3(\murec,\musen,\mucmp)\frac{1}{n}\ln n\\
+h_4(\murec)\ln n+h_5(\thetarec), 
\end{multline}
where $h_5(\thetarec)= 4I_{ov}\thetarec$. Recall the other notations have been defined in Appendix \ref{proof_convex}. 
In the following proofs, to provide a clear insight on comparison, we omit two terms in the expected latency for both methods: we omit $h_4(\murec)$, as it is negligible for high-complexity cases with $W_\textsc{O}\gg k$; we omit $\Tenc+\Tdec$, as it is minor compared with $\Tcmp_i$. With these terms omitted, let $\mathbb{E}[\Tc_m(n,k)]$ and $\mathbb{E}[\Tu_m(n)]$ denote the expected latency of coded and uncoded method, respectively. 

\subsubsection{Proof of Proposition \ref{prp:cp2uncoded}}
Based on \eqref{eq:tc-approx} and \eqref{eq:tu-approx}, comparing $\mathbb{E}[\Tc_m(n,k)]$ and $\mathbb{E}[\Tu_m(n)]$ is equivalent to comparing $R$ and $\max_k h(n,k)\triangleq{(k\ln n-n\ln{n}/{(n-k)})}{(n-k)}$.
Given $n$, $h(n,k)$ is maximized at $\partial h(n,k)/\partial k=0$, where the optimal splitting strategy is $k^*_{sub}(n)=n-e$, where $e$ is the natural base, and
$h(k^*_{sub}(n))={n}/{e}-\ln n$. Since $h(k^*_{sub}(n))$ is monotonically increasing in $n\geq10$ and $h(k^*_{sub}(10))=1.38>R$, we have $h(k^*_{sub}(n))>R$ for $n\geq 10$. 
Thus, $\mathbb{E}[\Tc_m(n,k^*_{sub})]<\mathbb{E}[\Tu_m(n)]$, which is $\Delta>0$.

\subsubsection{Proof of Proposition \ref{prop:fail}}
Based on Proposition \ref{prp:cp2uncoded}, there exists a $k^*_{sub}$ such that $\mathbb{E}[\Tc_m(n,k^*_{sub})]<\mathbb{E}[\Tu_m(n)]$ holds when there is no device failure. When one device fails,
$\mathbb{E}[\Tc_m(n,k^*_{sub})]$ becomes the $k^*_{sub}$-th order statistics of $n-1$ random variables, thus
is increased by $\frac{1}{k^*_{sub}}(\ln\frac{k^*_{sub}-1}{n-k^*_{sub}-1}-\ln\frac{n}{n-k^*_{sub}})$, which is less than $0.09$ for $n\geq 10$ . On the other hand, $\mathbb{E}[\Tu_m(n)]$ is increased by at least $R^{\textrm{cmp}}(n)>0.1$.
Therefore,
$\mathbb{E}[\Tu_m(n)]-\mathbb{E}[\Tc_m(n,k^*_{sub})]>\Delta$.

\subsection{Detailed Implementation of LtCoI}\label{LtCoI_detail}
We implement LtCoI using \verb|asyncio| in python. To begin with, a type-1 convolution task is split into $k$ pieces  (where $k$ can be larger than $n$), each is known as a \textit{source symbol}. Then, for the encoding process, LT codes generate \textit{encoded symbols} continuously, corresponding to the rateless property. Each time a random degree $d$ is sampled from Robust Soliton distribution \cite{Mallick2022rateless}, then $d$ \textit{source symbols} are selected uniformly and summed up to generate an encoded symbol, and the corresponding \textit{encoding vector} has a length of $k$ with 1's at the selected indices of the \textit{source symbols} and 0's elsewhere.

For each coded computation, the master creates $n$ coroutines to distribute subtasks to $n$ workers, and starts receiving the encoded outputs and the respective \textit{encoding vectors}. In each coroutine, encoded symbols are continuously created and sent to the worker. Upon receiving subtasks, the workers simply execute the received subtasks and send back the results. As for decoding, since the required number of encoded outputs $n_d$ for decoding is a random variable, we use Gaussian Elimination to judge the completeness of the received results. Once the rank of the \textit{encoding matrix} formed by encoding vectors of the received results is $k$, $n_d$ is determined and $\textbf{O}$ can be resolved.

\end{document}